\documentclass[journal,12pt,onecolumn,draftclsnofoot]{IEEEtran}

\usepackage{cite}

\ifCLASSINFOpdf
\else
\fi

\usepackage[inline]{enumitem}
\usepackage{amsmath}
\usepackage[cmintegrals]{newtxmath}
\usepackage{mathtools}
\usepackage{interval}
\usepackage{amssymb}
\usepackage{bbm}
\usepackage{algorithm}
\usepackage{algpseudocode}
\usepackage[short]{optidef}
\usepackage{graphicx}
\usepackage{subcaption}
\usepackage{balance}
\usepackage{microtype}
\usepackage{url}
\usepackage{xcolor}
\usepackage{color,soul}

\usepackage[T1]{fontenc} 

\makeatletter
\newcommand{\bdiv}{\mathbin {\operator@font div}}
\makeatother

\makeatletter
\def\BState{\State\hskip-\ALG@thistlm}
\makeatother

\let\downto=\downarrow
\makeatletter
   \newcommand{\argmin}{\mathop {\operator@font arg\,min}}
   \newcommand{\units}[1]{[{\operator@font #1}]}

   \def\minmaxtw@#1#2#3{#1\left\{#2,\;#3\right\}}
     \def\mintwo{\minmaxtw@\min}
     \def\maxtwo{\minmaxtw@\max}
\makeatother

\newcommand{\iris}{\textsf{\small{\mbox{Iris}}}}
\newcommand{\iriss}{\textsf{\small{\mbox{Iris}}} }

\begin{document}

\title{Iris: Deep Reinforcement Learning Driven Shared Spectrum Access Architecture for Indoor Neutral-Host Small Cells}

\author{Xenofon~Foukas,~\IEEEmembership{Member,~IEEE,}
        Mahesh~K.~Marina,~\IEEEmembership{Senior Member,~IEEE}
        and~Kimon~Kontovasilis}


\maketitle

\begin{abstract}
We consider indoor mobile access, a vital use case for current and future mobile networks. For this key use case, we outline a vision that combines a neutral-host based shared small-cell infrastructure with a common pool of spectrum for dynamic sharing as a way forward to proliferate indoor small-cell deployments and open up the mobile operator ecosystem. Towards this vision, we focus on the challenges pertaining to managing access to shared spectrum (e.g., 3.5GHz US CBRS spectrum). We propose \iris, a practical shared spectrum access architecture for indoor neutral-host small-cells. At the core of \iriss is a deep reinforcement learning based dynamic pricing mechanism that efficiently mediates access to shared spectrum for diverse operators in a way that provides incentives for operators and the neutral-host alike. We then present the \iriss system architecture that embeds this dynamic pricing mechanism alongside cloud-RAN and RAN slicing design principles in a practical neutral-host design tailored for the indoor small-cell environment. Using a prototype implementation of the \iriss system, we present extensive experimental evaluation results that not only offer insight into the \iriss dynamic pricing process and its superiority over alternative approaches but also demonstrate its deployment feasibility.
\end{abstract}


\section{Introduction}
\label{sec:intro}

\subsection{Background and Motivation}

\IEEEPARstart{M}{obile} data traffic growth over the past decade and forecasts have been driving research on scaling capacity of mobile networks. Much of this demand is from indoors, amounting to 80\% as of 2014 according to a Gartner study and expected to rise to over 95\% by the time 5G gets deployed~\cite{cisco-5g}. Indoor cellular coverage, however, has traditionally been poor. Outdoor solutions for indoor coverage are expensive due to building penetration losses~\cite{mason2016}. Even Distributed Antenna Systems (DAS) are found to be expensive except for a few large venues like stadiums~\cite{viavi-das, multi-carrier-2020}. Indoor small cells are considered relatively promising to address the coverage issue and scale the infrastructure with user density/demand. Indeed, making cells smaller and denser has historically been the biggest contributor to capacity scaling of cellular networks~\cite{zander13}. Despite this potential, indoor small cell deployments have been hampered due to operator concerns over deployment costs (and return on that investment) and issues such as site access and backhaul.\looseness=-1 

For the cost-efficient and simplified deployment of indoor small-cell networks for all operators, there is an emerging consensus around the notion of a {\em``neutral-host''}~\cite{multioperator_4g_americas,enterpriseiot-2018,linkedin-2018,denseair-2018,ipaccess-sumo-2018,baicells-2018,crown-castle-2018,giannoulakis2016emergence,matinmikko2017micro,ahokangas2016future}. The key idea is that the site owner (i.e. the neutral-host) offers indoor mobile access as a building amenity by taking the responsibility of deploying and managing the small-cell infrastructure and by allowing multiple operators to share it for a fee that covers the neutral-host's CapEx and OpEx (e.g. deployment, management and electricity cost), thus offering {\em small-cells as a service}\footnote{The neutral-host is a more general concept that could also be applicable in other settings (e.g. outdoor, rural, etc.), however our focus here is on the indoor scenario.}. The neutral-host becomes the only entity that needs to address issues such as power and backhaul, relieving the operators of deploying their own infrastructure and dealing with the associated challenges. Considering the ever-increasing significance of mobile access for users, the site owner is motivated to act as a neutral-host by the need to provide a high quality of experience for the building residents and visitors (thus increasing the value of the property), while the operators are motivated to pay a fee to gain indoor access rather than relying on their outdoor RAN infrastructure in order to improve their indoor coverage~\cite{ongo_white_paper, ahokangas2018business} (although providing a service with degraded quality to indoor users through their outdoor infrastructure is still a valid option to avoid paying a fee to the neutral-host).\looseness=-1 

As virtualization is a natural means for sharing the small-cell infrastructure, the neutral-host concept aligns well with the 5G vision of supporting a diverse array of services across different Mobile Virtual Network Operators (MVNOs) and verticals via network slicing. From this perspective, the neutral-host provides each operator (also referred to as a {\em tenant} henceforth) a virtual radio access network (vRAN) spanning the area of the indoor environment it covers; this vRAN becomes part of the operator's end-to-end network solution, including its existing core network or a cloud realization of the core (e.g., \cite{echo-mobicom18}). However, a vanilla realization of the neutral-host concept that serves just traditional mobile network operators (MNOs) bringing their own licensed spectrum offers limited incentives for the neutral-host and operators alike~\cite{multioperator_4g_americas}.

We envision that the potential of the neutral-host's infrastructure sharing capability would be significantly amplified through access to a pool of spectrum that is dynamically shared among operators. Firstly, traditional MNOs would be able to gain access to additional spectrum for increasing their capacity and for offloading purposes. Secondly, by removing the requirement to possess licensed spectrum (which typically only a handful of operators have), it allows new non-traditional operators into the fray, who may come with innovative revenue models differing from the traditional subscription-based model (e.g., free access that is monetized by advertising and analytics {\em a la} Internet services and free mobile apps). Lastly, the aforementioned increase in the network capacity offered by the additional spectrum and the potential cost reductions offered to users (or even free access) can greatly improve their quality of experience, which as already mentioned is the main incentive of the site owner to act as a neutral-host in the first place. In fact, there is more support for the neutral-host model following the 3GPP defined multi-operator core network (MOCN) form of network sharing, which requires use of common spectrum shared between operators~\cite{multioperator_4g_americas}.\looseness=-1 

The neutral-host's common (and dynamic) spectrum pool could in principle be made up of licensed spectrum pooled from different MNOs, unlicensed spectrum or shared access spectrum~\cite{atis2016}. Regarding the latter, recent regulatory developments below 6 GHz allow sharing of lightly used spectrum held by legacy or public-sector incumbents (e.g., radars) via tiered spectrum access models~\cite{spectrum-2018, fcc2015, ofcom2016}, offering substantial amounts of spectrum at a lower acquisition cost compared to licensed spectrum and without the complex coexistence issues of unlicensed spectrum. The Citizen Broadband Radio Service (CBRS) in the US~\cite{cbrs_white_paper} is a case in point, allowing the shared use of the 3.5 GHz band via a three-tier access model; a management entity called Spectrum Access System (SAS) ensures that when higher tier users need to use the spectrum, they get interference protection from lower tier ones. In fact, a new LTE-based service called OnGo that is operating over CBRS spectrum is promoted by the CBRS Alliance as an ideal solution for neutral-host deployments~\cite{ongo_white_paper}. Licensed shared access (LSA) model for spectrum sharing~\cite{matinmikko2014spectrum} that is promoted for some bands in Europe, especially in its dynamic form, is another such relevant development. {\em In view of the above, we consider the scenario where the neutral host is powered by shared access spectrum in the style of CBRS or LSA.} \looseness=-1

\subsection{Paper Overview and Contributions}
\label{sec:dynamic_pricing_goals}

The focus of this work is on addressing the challenges that arise with respect to managing access to shared spectrum in an indoor neutral-host small-cell environment, which constitute {\bf the requirements for the desired system}: 
\begin{enumerate}

\item As the neutral-host needs to support multiple (traditional and non-traditional) operators all competing in offering broadly the same type of service to users, the system should facilitate for each tenant service differentiation over rival tenants and control over its share of resources without requiring direct/explicit interaction among tenants. These are key concerns from the operators perspective to incentivize their participation in neutral-host small cells~\cite{multioperator_4g_americas}. This requirement means that tenants should not have to reveal any private information regarding their business model to the neutral-host; instead they should operate in isolation with respect to each other and the neutral-host, and should be able to dynamically change their private spectrum valuations.

\item The system should provide a control mechanism to enable the efficient and dynamic spectrum sharing among tenants by aiming to closely match the spectrum supply with the tenants' demand. This should be done in a way that tenants who value the spectrum most get it, especially during periods of congestion (e.g., due to insufficient spectrum availability). The allocation of radio resources to tenants should be performed in a way that respects this constraint. Being in a position to satisfy the tenants' demand, also implies the satisfaction of their service level agreements (SLAs). However, given that the shared spectrum availability can change dynamically and unexpectedly over time, hard SLAs may be infeasible for the tenants and therefore the desired system should aim to provide soft SLAs instead.
\item The neutral-host should be able to cover its expenses for offering the service, including the fixed costs (e.g. deployment, management and electricity) and also a time-varying spectrum acquisition cost~\cite{matinmikkoblue2018analysis}, depending on the amount of the shared spectrum acquired to meet the overall demand. This last cost needs to be recouped from the tenants in a dynamic manner, since any pre-agreed static fee may either overcharge the tenants or put the neutral-host in losses. Crucially, as already explained, the primary goal of the neutral-host is to provide a high quality of experience for building residents and visitors, and doing so without incurring losses. So revenue maximization is not the main driver although a revenue target linked with the neutral-host's incentive to provide the service with some profit margin (adjusted depending on the deployment environment) could also exist. Note that the environment in which the neutral-host operates is not a monopoly (e.g., tenants could opt to use their own external RAN with degraded quality of service). 
\item The solution approach meeting the above requirements should be realizable in the context of a shared spectrum based neutral-host small cell system architecture that is practical in terms of algorithmic complexity, signaling overhead, etc. 

\end{enumerate}

Our key insight in this paper is that pricing can be an effective control mechanism to meet the first two aforementioned requirements. Pricing has been effectively employed in other contexts~\cite{kelly1998rate,sen2015smart,li2011optimal,liu2014pricing} to regulate demand and enable efficient sharing of resources with service differentiation, while it also naturally allows meeting the third requirement of neutral-host cost recovery and achieving a revenue goal if it exists. Given that tenant behaviors and traffic demands as well as spectrum availability can vary over time, a single optimal fixed price may not exist and thus pricing has to be dynamic. On the complementary side, we view a cloud RAN (C-RAN)~\cite{chih2014recent} architecture to be more suitable for the indoor neutral-host small cell environment, due to the better scaling it offers in terms of spectrum availability, number of tenants etc., while allowing a cheaper and denser small-cell radio infrastructure. The result is our proposed approach \iris, a novel dynamic pricing shared spectrum access architecture for indoor neutral-host small-cells. The key components of the proposed \iriss approach and our contributions are outlined below:\looseness=-1

\begin{itemize}
\item (\S\ref{sec:multi_design}) We design a neutral-host system following C-RAN and RAN slicing design principles that embeds a dynamic pricing mechanism that regulates the allocation of spectrum to tenants by determining the price at which tenants can obtain a share of spectrum at any given time instant, while also considering the cost/revenue requirements of the neutral-host.

\item (\S\ref{sec:multi_model}) In view of the stochastic nature of the neutral-host's environment with several unknowns (tenant behaviors, future demands and spectrum availability), we model the pricing decision problem as a Markov Decision Process (MDP) and resolve it using reinforcement learning. As the large state space and continuous action space of the problem make common reinforcement learning techniques slow and inefficient (as we experimentally demonstrate), we leverage recent machine learning advances and employ deep reinforcement learning. Unlike the \iriss dynamic pricing mechanism, existing spectrum sharing mechanisms relevant for the neutral-host context~\cite{sciancalepore2017mobile,caballero2017multi,bega2017optimising,kokku2012nvs,kokku2013cellslice,jiang2016network,crippa2017resource,kibria2017shared} fail to meet some of the necessary requirements listed above, as discussed in the next section. 

\item (\S\ref{sec:impl}) We develop a prototype implementation of \iris, with the goal of demonstrating the feasibility of our proposed mechanism, thereby satisfying the fourth practicality requirement above. To our knowledge, relative to existing neutral-host designs~\cite{fajardo2016introducing,chochliouros2017novel,foukas2017orion}, this is the first design accounting for the peculiarities of spectrum sharing and the indoor small cell environment, along with a concrete implementation. \looseness=-1

\item (\S\ref{sec:multi_eval}) Using the above mentioned prototype implementation, we demonstrate the system's feasibility in practice and conduct extensive experimental evaluations --- characterizing the learning behavior of \iris, its performance in different conditions, and highlighting its superiority with respect to static pricing and alternative approaches from the literature~\cite{low1999optimization,kibria2017shared}.

\end{itemize}

\section{Related Work}
\label{sec:related}
\noindent\textbf{Dynamic pricing in other contexts.} Fundamentally, the pricing problem we have bears similarity with the pricing work in the Internet congestion control context~\cite{kelly1998rate,low1999optimization}. In these works, pricing is used as a signal to regulate the rates of senders for efficiently sharing network resources (e.g., bandwidth of links). Referring to \cite{low1999optimization}, for example, each link in the network sets a price depending on its aggregate demand from all senders and each sender adjusts its rate independently in a way that maximizes its net utility after accounting for the bandwidth cost. The key difference from our case is that these works do not have the equivalent of requirement (3) (\S\ref{sec:dynamic_pricing_goals}), regarding the need of the neutral-host to reach a revenue target that will allow it to cover its expenses. 

Dynamic pricing has also been successfully applied in various other contexts where regulation of demand is required~\cite{ha2012tube,li2011optimal,liu2014pricing}. In those cases the focus is on controlling the end-user demand by the operators, unlike our case of spectrum sharing among operators via the neutral-host.  

\noindent\textbf{Spectrum sharing in the RAN slicing context.} Neutral-host spectrum sharing can be seen as a specific form of RAN slicing and as such, RAN slicing mechanisms are relevant. There exist several algorithmic works~\cite{sciancalepore2017mobile,caballero2017multi,bega2017optimising,kokku2012nvs,kokku2013cellslice,jiang2016network,crippa2017resource} focusing on either the base station level (e.g., \cite{kokku2012nvs,jiang2016network}) or the RAN level (e.g., \cite{kokku2013cellslice,caballero2017multi,sciancalepore2017mobile}), allocating radio resources to slices based on their SLAs. As all these mechanisms centralize the resource allocation at the infrastructure provider (neutral-host in our setting), they fail to meet requirement (1) (\S\ref{sec:dynamic_pricing_goals}). Also, with the exception of \cite{bega2017optimising} where revenue maximization for the infrastructure provider is considered, others do not meet requirement (3) of recovering costs and reaching the revenue target of the neutral-host. With respect to requirement (2), the focus on strict SLAs in these works may also be limiting when dealing with shared access spectrum.\looseness=-1

A recent work~\cite{kibria2017shared} explicitly targets the shared spectrum  neutral-host setting but shares the same limitations as the above mentioned works. It presents several pre-determined spectrum allocation policies at the neutral-host, mostly SLA based with the exception of one that assumes all tenants have the same utilities and allocates spectrum proportional to their traffic loads. We consider the latter in our comparative evaluations to highlight the service differentiation benefit of \iris. 

\noindent\textbf{Spectrum sharing without infrastructure sharing.}
The allocation of shared spectrum has also been considered in settings where operators deploy independent infrastructures~\cite{kibria2016resource,kibria2017heterogeneous,luo2014multi,sanguanpuak2017spectrum,singh2015co,singh2015repeated,hasan2018communication}. Some works assume that participating operators have predetermined agreements regarding their priority for accessing the spectrum in cases of congestion~(e.g., \cite{kibria2016resource}), while others focus on the architectural aspect of the system~(e.g., \cite{luo2014multi}) or on the coordination among operators (e.g., ~\cite{singh2015repeated,singh2015co}).  

\noindent\textbf{Auction-based dynamic spectrum sharing mechanisms.} A rich body of literature on dynamic spectrum auction mechanisms is broadly related~\cite{zhou2008ebay,fu2013stochastic,zhu2016virtualization,feng2015flexauc,le2014new,sengupta2008designing,gandhi2008towards,jia2009revenue,gao2013integrated,dong2016double}. The most relevant from our context are \cite{fu2013stochastic,zhu2016virtualization} but both have limitations from a practicality standpoint. The mechanism in \cite{fu2013stochastic} requires continual exchange of information between tenants and the neutral-host about each end-user device, and it allows only discrete number of traffic rates for tenant resource requests. \cite{zhu2016virtualization} proposes a hierarchical auction-based mechanism that requires the involvement of end-users in the auction. 

More fundamentally, any auctioning mechanism for our setting has to handle a time-varying spectrum acquisition cost for the neutral-host along with its revenue target, which requires a {\em dynamic} reserve price. Setting such a reserve price statically and in a naive way effectively leads to the same limitation as in the case of~\cite{low1999optimization} in terms of meeting requirement (3) (something that is also demonstrated in Section~\ref{sec:eval_alt}). In contrast, in this work we demonstrate the capability of \iriss to meet the aforementioned requirement and to effectively satisfy the tenants' demand through the system's dynamic pricing decisions. Furthermore, it should be noted that \iris's decisions could also be used as an enabler of auction-based spectrum sharing mechanisms for our setting in cases where price differentiation among tenants is required, i.e. by setting the reserve prices of an auction-based scheme based on the pricing decision of \iris.

\noindent\textbf{Neutral-host system designs and specifications.} A number of recent designs that consider multi-tenancy support in mobile RANs are applicable to the indoor neutral-host small cell setting. Perhaps the ones most relevant are: Orion~\cite{foukas2017orion}, SESAME~\cite{fajardo2016introducing} and ESSENCE~\cite{chochliouros2017novel}. However, these works do not consider the use of shared spectrum and its implications, the main focus of this paper.\looseness=-1

In terms of specifications targeting the neutral-host setting, nFAPI~\cite{nfapi_specs} is the most relevant one in which a functional split at the MAC layer is specified and each virtual operator is assigned a VNF implementing the higher-layer protocols. However, in contrast to our work, each tenant is assigned a static chunk of spectrum. Another closely related specification is MulteFire~\cite{alliance2017multefire}, which is a form of LTE deployment in unlicensed bands. In contrast to our work, the focus of MulteFire is on the ways to enable co-existence with other technologies operating over unlicensed spectrum (e.g. Wi-Fi). \looseness=-1 

\section{Iris System Architecture}
\label{sec:multi_design}

\begin{figure}[t]
	\centering
	\includegraphics[width=0.7\columnwidth]{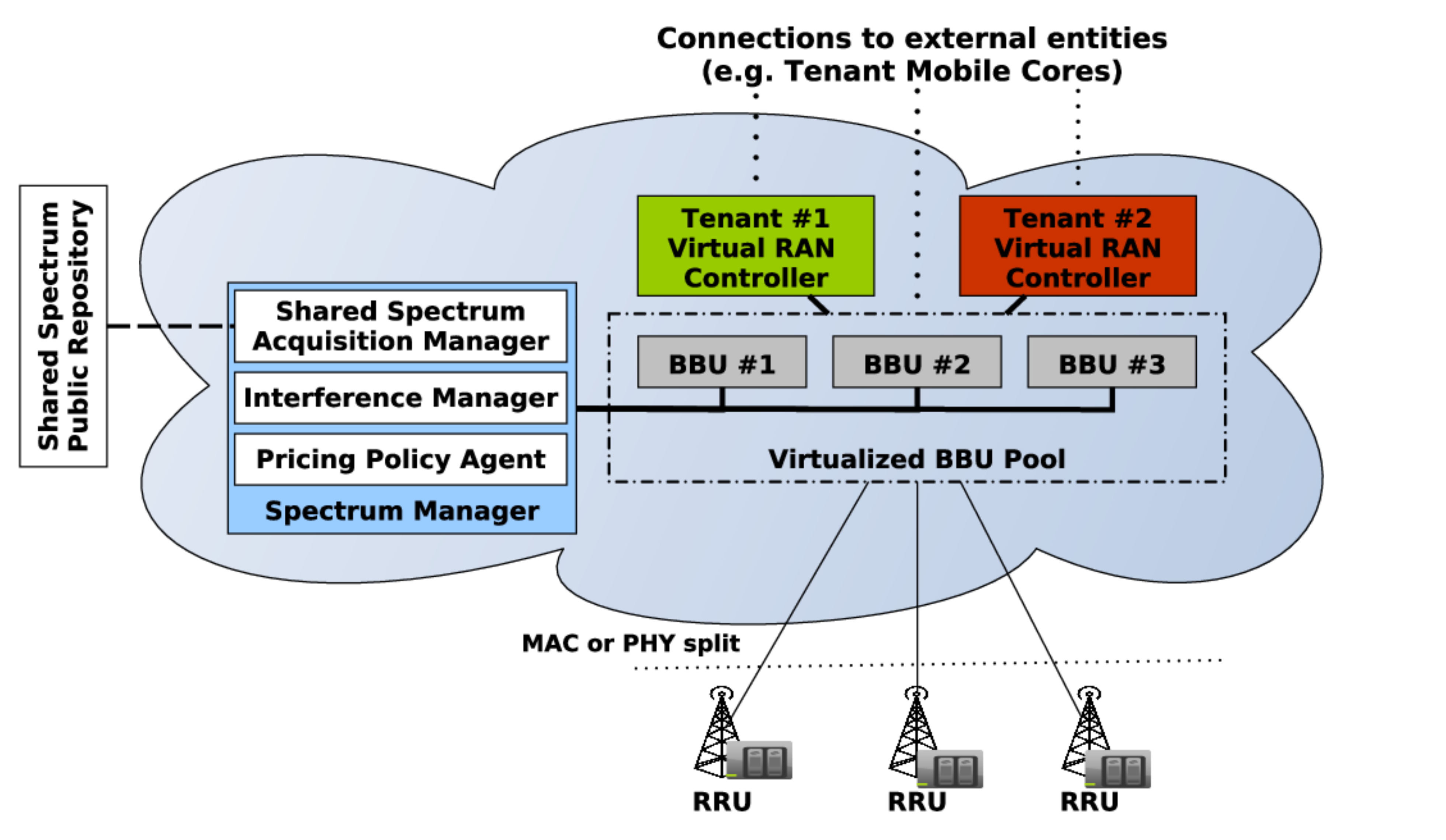}
	\caption{Schematic of \iriss neutral-host system architecture.}
	\label{fig:high_level_multi}
	\vspace{-2mm}
\end{figure}
The design of \iriss builds on the observation that the small-cell infrastructure sharing capability offered by a neutral-host is a particular albeit compelling use case of the broader RAN slicing in the 5G context. However, a vanilla RAN slicing system would be insufficient to address the specific needs of shared spectrum management and indoor small-cell environments. The design of \iriss addresses these needs (Fig.~\ref{fig:high_level_multi}) by embracing the cloud RAN (C-RAN) paradigm, with baseband processing units (BBUs) centralized in a virtualized BBU pool located in an edge cloud (e.g., in the basement of the indoor space) and remote radio units (RRUs) deployed throughout the building in a planned manner. The RRUs are connected to the BBUs over high speed channels (e.g., 10-Gigabit Ethernet or Fiber). This approach places most of the RAN processing on the edge cloud which allows the system to scale better as BBU resources can be adaptively provisioned depending on the number of tenants and the spectrum availability. It also lowers the form factor of the RRUs, making their deployment easier and discreet from a building aesthetics viewpoint.\looseness=-1 

Each tenant is allocated a \emph{Virtual RAN Controller}, deployed as a Virtual Network Function (VNF) over the edge cloud. The controllers interface with the BBUs using message-based communication and provide tenant-specific functions such as schedulers and mobility managers, as well as an agent for the allocation of shared spectrum (discussed shortly).

At the heart of \iriss lies the \emph{spectrum manager}, a centralized controller managed by the neutral-host. This controller informs the BBUs about the amount and type of available spectrum (shared or privately owned) and about its valid allocations, depending on the access rights of tenants, distinguishing in particular between tenants operating exclusively over shared spectrum and tenants that can also use their own private licensed spectrum. A shared spectrum acquisition manager acquires the shared spectrum in a demand driven manner through a public repository (e.g., SAS in the CBRS context). Moreover, this controller manages interference among small cells. Due to the system's C-RAN based design, the VNF of the spectrum manager co-exists with the virtualized BBU pool over the same edge cloud, simplifying its coordination with the BBUs through low-latency and high bandwidth channels and enabling the use of advanced interference management techniques like Coordinated Multipoint (CoMP)~\cite{irmer2011coordinated}. 

\begin{figure}[h]
	\centering
	\includegraphics[width=0.7\columnwidth]{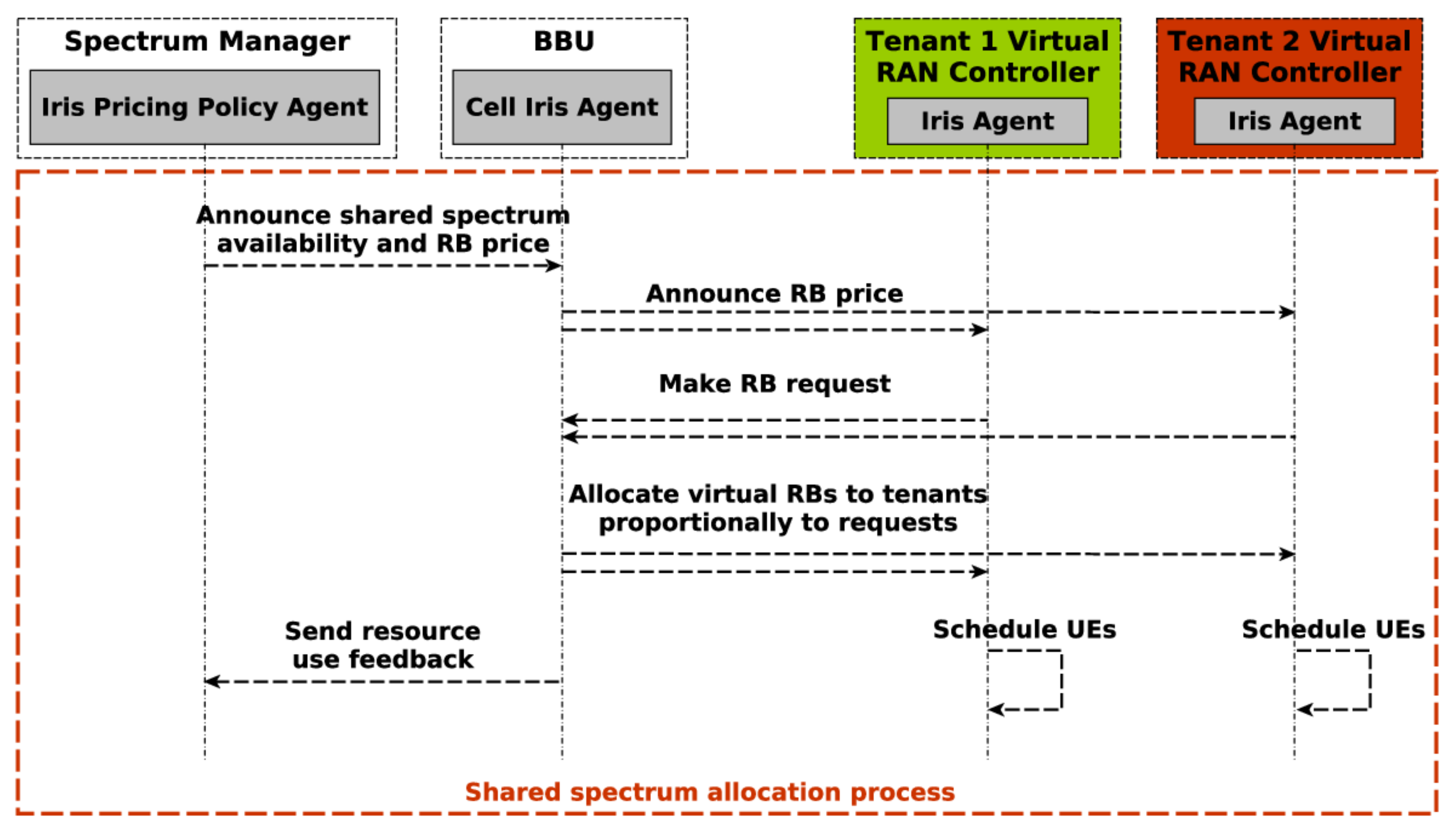}
	\caption{\iriss agents involved in dynamic pricing mechanism.}
	\label{fig:dynamic_pricing_multi}
	\vspace{-2mm}
\end{figure}

\noindent{\bf Shared spectrum allocation process in \iris}. Crucially, the spectrum manager hosts a pricing policy agent and is responsible for dynamically deciding the price for the tenants to use shared spectrum resources. The dynamic pricing mechanism of \iriss follows a time slotted operation for the allocation of shared spectrum, with each slot referred to as an \emph{epoch} henceforth. The functionality of the dynamic pricing mechanism
is distributed among three distinct agents as illustrated in Fig.~\ref{fig:dynamic_pricing_multi}.\looseness=-1

The \emph{pricing policy agent} initiates the shared spectrum allocation process in each epoch, deciding on the price for each cell using a deep reinforcement learning algorithm that is described in depth in \S\ref{sec:ddpg}. The pricing policy agent announces the current epoch, the spectrum availability and the cell specific prices to the respective \emph{cell agents} residing in the BBUs, which in turn convey the price to the \emph{tenant agents} residing in the tenants' virtual RAN controllers. Each tenant considers the announced price along with its traffic load at the cell in question to decide on quantity of resource to be requested as dictated by its internal {\em private} policy. The tenant requests are aggregated at the cell agent, which distributes the available shared spectrum proportionally to the tenants' requests and notifies the pricing policy agent about the allocated resources, the load of the tenants etc. The schedulers running in the virtual RAN controllers of the tenants use the allocated resources to serve the traffic of their UEs as per the tenants' internal policies. Once the allocation process is complete, the pricing policy agent uses the feedback obtained from the cells in terms of the behavior of the tenants and decides on a new price for the upcoming epoch. \looseness=-1 

\section{Iris Dynamic Pricing Mechanism}
\label{sec:multi_model}

In this section, we describe the core component of \iriss -- its dynamic pricing mechanism. 

\subsection{System Model}
\label{sec:neutral_host_model}

{\bf Tenant resource requests.} In our model, tenants express their resource requests in terms of radio resource blocks (RBs) through the \iriss tenant agents. We assume that tenants have a way to map their aggregate throughput demands into the number of RBs required (e.g., by assuming an average spectral efficiency for every RB). Such a mapping is reasonable, considering that indoor small cell deployments are typically planned to provide near-optimal performance to users (UEs) within 20-30m~\cite{salami2013lte}.

{\bf Neutral-host and tenant interaction.} As already described, the neutral-host follows a time slotted operation in the form of epochs for the allocation of shared spectrum. The duration of an epoch is expected to be short (e.g., 20-100ms), allowing the neutral-host to allocate radio resources in real-time. In each epoch $t$, the neutral-host determines a resource block price $p^t~\in~\interval{p_{min}}{p_{max}}$ through its pricing policy agent with which the tenants can buy the available resources. Our model does not assume predetermined bounds to the range of prices, providing enough flexibility to set $p_{min}$ and $p_{max}$ according to the pricing characteristics of the specific domain in which \iriss is deployed. The range of possible prices is assumed to be known to the tenants a priori (e.g., specified in their contract), allowing them to develop their radio resource acquisition policies according to the expected prices. Without loss of generality, we consider dynamic pricing for the allocation of downlink radio resources; the uplink can be treated similarly. 

In each epoch $t$, all tenants see the price $p^t$ announced by the neutral-host (through the cell agents) and decide how many resources to buy. The neutral-host is oblivious to the behavior of the tenants, not knowing the internal mechanism (possibly changing over time) that governs their decisions. Consequently, the high level goal of the neutral host would be to ``predict'' the demand of tenants at any point in time and dynamically decide on a price that would utilize the resources as efficiently as possible while recovering its cost and maximizing the tenant satisfaction, by allocating the radio resources according to the expressed tenant demands. The model presented here is compatible with very general tenant behavior patterns, deterministic (e.g., driven by the optimization of utility functions) or not. Due to this generality, our model does not require a concrete definition of the tenants' behavior and simply relies on the fact that tenants express their demands in terms of resource block requests. It is noted that in Section~\ref{sec:multi_eval} and for the sake of concreteness of our evaluation, we model the different tenant behaviors in the form of a rich set of utility functions, which are used to demonstrate the effectiveness of \iris.\looseness=-1

To model the temporal evolution of the tenants' demand, we divide a day into $H$ periods, each $e$~epochs long, so that $He$~epochs make up 24h in the day. This construction makes a period correspond to an appropriate time interval within a day (e.g., an hour in a day) so that tenants' behavior is not expected to vary within a period but could across periods. Clearly, the shorter the period, the finer the granularity at which tenant behavioral changes can be captured. Without restricting generality, one may index periods within a day in the range $0\leq h<H$ and may take the evolution of the system to start at epoch $t=0$ coinciding with the beginning of a day. With this convention, the index of the current epoch~$t$ maps to the index of the current period of the day as: $h(t) = (t \bdiv e) \bmod H$. It should be noted that the scheme imposes a natural synchronization, in which all tenants can always refer to the correct current epoch. In the rest of this section, we will use superscripts of the form $\cdot^t$ to denote the time dependency of any quantity including cases when it occurs indirectly through~$h(t)$. 

{\bf Shared spectrum acquisition cost and revenue target of the neutral-host.} Let $n^t \in \mathbb{Z}$ be the number of RBs obtained by the neutral-host from an external/public spectrum repository in epoch $t$. To maintain flexibility, the pricing mechanism regards the interaction between the public repository and the neutral-host in abstract terms. Consequently, $\{\,n^t, t\ge0\,\}$ is a stochastic process and the neutral-host, although informed about the current value~$n^t$, is unaware of the process' future dynamics so dynamics of a very general form can be accommodated. The only assumption (to enable the MDP framework discussed later) is that $n^{t+1}$, conditioned on the value of~$n^t$, follows a probability distribution (unknown to the neutral-host) that may depend on~$t$ and/or the current load of the tenants. This is a very mild assumption compatible with virtually all scenarios of practical interest. 

To capture the neutral-host's incentive for participation, we introduce a target revenue level $T$. The value of $T$ represents the per epoch revenue that the neutral-host expects to obtain through the dynamic pricing scheme for the particular small-cell in question. Generally, $T$~can change dynamically as the neutral host seeks to offset its OpEx that encapsulates not just fixed costs like electricity and management of the infrastructure, but also dynamic costs like the the cost for the amount of RBs~$n^t$ obtained from the external spectrum repository in epoch~$t$. In the following we will use the notation $T(n^t)$ to signify this functional dependence. This notion of a target revenue level is general enough to also capture other types of expenses (e.g. electricity) and could also be used to accommodate more general profit aims of the neutral-host (e.g., to dynamically adjust its profit margin according to the conditions of its environment). 

{\bf System dynamics and neutral-host's small-cell resource allocation.} Let~$I$ be the set of tenants served by the system. For each tenant $i\in I$, the expected load of a cell in epoch $t$ is denoted by $l^t_i$, representing the total traffic that tenant~$i$ is expected to serve during epoch $t$. For example, this could be the bytes stored in the transmission buffers of all the UEs of the tenant in the cell and a forecast of any new traffic expected during epoch $t$. This can accommodate very general dynamics for the evolution of~$l^t_i$, for all $i\in I$. The only assumption made (to enable the MDP formulation) is that $l_i^{t+1}$, conditioned on the value of~$l^t_i$, follows a probability distribution that may depend on one or more of: the time~$t$, the amount of radio resources $n^t$, and the price $p^t$. \looseness=-1

The tenant~$i$'s behavior at epoch~$t$ is captured through its RB request~$\nu^t_i\in\mathbb{Z}$. The dynamics of~$\nu_i^t$ can be general, the only restriction being that $\nu^{t+1}_i$, conditioned on the value of  $\nu^t_i$, follows a probability distribution (unknown to the neutral-host), whose form may depend on one or more of: the time~$t$, the current tenant load~$l^t_i$ and the price~$p_t$. \looseness=-1

The collective request across all tenants may exceed the amount of available resources, i.e., it is possible for $\sum_{i\in I} \nu_i^t > n^t$. In such a case, the neutral-host would allocate the available resources proportionally to the tenants' requests. By using~$u^t_i$ to denote the resources actually allocated to tenant~$i$ in epoch~$t$, this allocation rule translates into
\begin{equation}
\label{eq:allocated_rbs}
u^t_i(\vec{\nu}^t)
	= \frac{\nu^t_i}{\sum_{j\in I} \nu^t_j}
	\min\{\,n^t, \sum_{j\in I} \nu^t_j\,\},
	\qquad\forall i\in I,
\end{equation}
where $\vec{\nu}^t$ stands for a vector containing the tenants' requests. Always, $\sum_{i\in I} u^t_i\leq n^t$. And $u_i^t=\nu_i^t$ for all $i\in I$, when $n^t \geq \sum_{i\in I} \nu_i^t$.

\subsection{Problem Formulation}
\label{sec:form_multi}

Given the system model just described, we now formulate the neutral-host's dynamic pricing problem as a discrete-time, continuous state and action space MDP. Essentially, the neutral-host observes the state of its environment and makes a decision for an action (price $p^t$) based on this observation, getting a reward while transitioning the environment into a new state. We denote this action as $a^t \in A$ where $a^t=p^t$.

The state ($x^t$) of the neutral-host's environment in epoch $t$ is made up of the vector $\vec{\nu}^{t-1}$ of virtual radio resources requested by the tenants in the previous epoch $t-1$, a vector $\vec{l}^t$ containing the current loads $l_i^t$  of the tenants and the number of available radio resources at the neutral-host $n^t$. That is, 
\begin{equation}
\label{eq:nh-state}
x^t := (\vec{\nu}^{t-1}, \vec{l}^t, n^t) \in X.
\end{equation}
Note that $x^t$ is known to the neutral-host as it either contains information maintained by itself or obtained from the tenants every time they request resources.

The reward function of the neutral-host is designed so that it can capture the first three requirements of the system as identified in Section~\ref{sec:dynamic_pricing_goals}. Using the action notation $a^t$ for the price decision $p^t$, the reward function is defined as

\begin{equation}
\label{eq:neutral_host_reward}
r(x^{t+1}, a^t \mid x^t) = f(\tfrac{n^t - \sum_{i \in I}\nu_i^t}{n^t})g(\tfrac{a^t\sum_{i \in I}u_i^t(\vec{\nu_i^t})}{T(n^t)}).
\end{equation}
Indeed, the first requirement of the system is met, as the tenant behaviors are hidden from the neutral-host, which only has tenant resource requests ($\nu_i^t$) to glean that information (as part of the state given as input to the reward function).

The second requirement, i.e., avoiding a mismatch between resource supply and demand is handled through the first factor on the right hand side of~(3). In the argument of the function $f$, this mismatch is expressed in a relative sense to make the reward function behave in the same way regardless of the amount of available spectrum. A zero value of this argument signifies a desirable perfect match between supply and demand. Positive values of the argument signify resource under-utilization ($n^t > \sum_{i \in I}\nu_i^t$).
Avoiding it would leave room to allocate more resources to tenants and increase their satisfaction. Negative values of the argument signify excess demand ($n^t < \sum_{i \in I}\nu_i^t$). This should be avoided, as the proportional allocation in rule~\eqref{eq:allocated_rbs} gives some tenants fewer RBs than those requested at this price, leading to a decrease in their satisfaction. These are all met by defining the function~$f$ as \looseness=-1
\begin{equation}
\label{eq:gaussian}
f(x) =  e^{-x^2/\sigma^2},\qquad \sigma>0.
\end{equation}
With this Gaussian form, the highest contribution to the reward is attained when $x=0$ (perfect match) while positive or negative mismatches are penalized at an exponential rate. The parameter $\sigma$ tunes the ``sensitivity'' of~$f$ -- smaller values of~$\sigma$ penalize resource mismatches more aggressively.

The second factor of~\eqref{eq:neutral_host_reward} corresponds to the third system requirement of avoiding a mismatch between the actual and target levels of revenue. The argument of the function $g$ is the ratio of these two levels. The ideal case is when this argument is equal to~1, a perfect match between actual and target revenues. When the argument is smaller than~1, the actual revenue is below the target. This should be avoided as it signifies a reduced incentive for the neutral-host to provide the service (the neutral-host is experiencing losses). When the argument is greater than~1, the revenue exceeds the target level. This should also be avoided as it suggests that a lower price could also satisfy the goal of the neutral-host which could potentially improve the satisfaction of the tenants. All these features can be incorporated, by letting
\begin{equation}
\label{eq:scale_factor}
g(x) = \Bigl(\frac{\min\{1,x\}}{\max\{1,x\}}\Bigr)^\delta, \qquad \delta \geq 0.
\end{equation}
The highest reward contribution occurs when there is a perfect match ($x=1$) while mismatches are penalized according to a power law. The value of $\delta$ tunes the sensitivity of~$g$, higher values of~$\delta$ penalizing revenue mismatches more aggressively. Moreover, the joint tuning of the parameters $\sigma$
and~$\delta$ can adjust the relative importance between the two factors $f$ and $g$ of the reward function. By making the values of any of these parameters smaller, the neutral-host tends to care more about the utilization of resources by the tenants and to disregard its own level of revenue. Increasing the values of the parameters has a reciprocal effect.

With the reward function~\eqref{eq:neutral_host_reward}, the behavior of the neutral-host is defined by a policy $\pi$, which maps the states to a probability distribution over the actions $\pi: X \rightarrow Pr(A) $. With the mild assumptions stated in \S\ref{sec:neutral_host_model} and the neutral-host's state as in~\eqref{eq:nh-state}, the state transitions from $x^t$ to $x^{t+1}$ given the action $a^t$ satisfy the Markov property and thus, applying a policy $\pi$ to this MDP defines a Markov chain. We denote expectations over this chain by $\mathbb{E}_\pi$. We define the return from a state $x^t$ as the sum of the discounted future rewards: $R^t=\sum_{\tau=t}^{\infty}\gamma^{(\tau-t)}r(x^{\tau+1}, a^\tau \mid x^\tau)$ for a discounting factor $\gamma~\in~\interval{0}{1}$~\cite{puterman2014markov}. The goal of the neutral-host then is to find a pricing policy that will maximize its expected returns from the start state $\mathbb{E}_\pi\left[R^0\right]$ over an infinite horizon. It should be noted that the choice for using a discounted rather than an average reward was mainly driven by the unpredictability of the environment, which can potentially change over time (e.g. tenants changing their spectrum acquisition policy).

\subsection{Deep Reinforcement Learning Solution}
\label{sec:ddpg}

Reinforcement learning is a common way to solve MDP problems where an exact model describing the dynamics of the environment (e.g., tenant behavior and network traffic in our case) is unavailable. Q-learning~\cite{watkins1992q} is a well-known algorithm for such problems. Q-learning employs an action-value function $Q^\pi$ which describes the expected future return after taking the action $a^t$ in some state $x^t$ and following policy $\pi$ from that point on, i.e.,
\begin{equation}
	Q^\pi(x^t, a^t) = \mathbb{E}_\pi\left[R^t | x^t, a^t\right].
\end{equation}
This function can be expressed through a recursive relationship known as the Bellman equation:
\begin{equation}
	Q^\pi(x^t, a^t) = \mathbb{E}_{r^t,x^{t+1} \scriptstyle \sim X}\left[r(x^t,a^t) + \gamma Q^\pi(x^{t+1}, \pi(x^{t+1}))\right].
\end{equation}
The policy used for the estimation of the discounted future reward of Q-learning is the greedy policy $\pi(x) = \arg \max_a Q(x,a)$ whereas an exploration policy is employed for the state transitions (e.g., take random actions). This makes Q-learning an off-policy method in that the policy $\pi$ used to estimate the discounted future reward is different from the policy used for the action of the learning agent in a state transition.\looseness=-1

Though an obvious choice, Q-learning is not appropriate to our problem for several reasons. Firstly, it uses a table to store its Q-values. When the state space of the problem is continuous or very large (as in our problem due to the range of possible values for $l_i$, $\nu_i$ and $n$), calculating $Q^\pi$ using a table becomes challenging. To overcome this, we need to rely on function approximators~\cite{busoniu2010reinforcement} parametrized by $\theta^Q$. These parameters can be optimized by minimizing the loss:\looseness=-1
\begin{equation}
	L(\theta^Q) = \mathbb{E}_{\pi^\prime}\left[(Q(x^t,a^t|\theta^Q) - y^t)^2\right],
\end{equation} 
where
\begin{equation}
y^t = r(x^t, a^t) + \gamma Q(x^{t+1},\pi(x^{t+1}) | \theta^Q).
\end{equation}

In addition to the large state space, we also have to deal with a continuous action space (the announced price) which needs to be discretized in order to use Q-learning. However, there is no obvious or straightforward way to discretize the prices since the price range and its interpretation can be environment dependent~\cite{busoniu2010reinforcement}. 

We find that policy gradient actor-critic algorithms~(e.g., \cite{silver2014deterministic}) are more suitable for our purpose. Such algorithms maintain a parametrized actor function $\pi(x|\theta^\pi)$ that estimates an action policy and 
a parametrized critic function $Q(x,a | \theta^Q)$ that estimates the Q-values of action-state pairs through the Bellman equation, as in Q-learning. The actor policy is improved at each step by performing a gradient descent considering the estimated values of the critic. Recent works (e.g., \cite{lillicrap2015continuous,schulman2015high,duan2016benchmarking,gu2017deep}) show that using deep neural networks as the function approximators for the estimation of actors and critics can produce better results compared to using linear approximators, when the learning task presents similar complexity to the one we have in terms of its dimensionality, including higher rewards (avoiding local optima) and improved convergence speed in some cases. 

In view of the above, we choose to use a state-of-the-art deep reinforcement learning actor-critic algorithm called DDPG~\cite{lillicrap2015continuous}, which has been shown to consistently provide good results for a wide range of problems and learning environments~\cite{tassa2018deepmind, duan2016benchmarking}. The use of deep neural network approximators allows DDPG to scale to high-dimensional state spaces and operate over continuous action spaces, ideal for our problem. One of its key features is the use of replay buffers (a type of cache) to sample prior transitions $(x^t,a^t,x^{t+1},a^{t+1})$ which are used to train the neural networks. It also uses a technique called batch normalization that improves the effectiveness of the learning process when using features with different units and ranges (e.g., RBs and time). Finally, it uses a technique that employs slow-changing copies of the actor and critic networks, called target networks, which are used for calculating $y^t$. This has been shown to greatly improve the stability of the learning method.\looseness=-1

\begin{algorithm}
\small
	\caption{\iriss Dynamic Pricing Mechanism Outline}\label{euclid}
	\label{alg:ddpg}
	\begin{algorithmic}[1]
		\Procedure{DynamicPrice}{}
		\State $t \gets 0$
		\State Receive initial network state $x^0$ (\texttt{state in first epoch of the day})
		\BState \emph{loop}:
		\State Choose a price $p^t$ given the policy of the actor
		\State $a^t \gets p^t + \epsilon$, where $\epsilon$ is exploration noise
		\State Execute action $a^t$ (\texttt{announce price to tenants})
        \State Collect the radio resource requests of tenants $\vec{\nu^t}$ and distribute the allocated RBs $u_i^t(\vec{\nu^t}),~\quad \forall i \in I$ 
        \State Calculate the reward $r^t$ and transition to the state $x^{t+1}$
		\State Update the actor-critic parameters $\theta^Q$ and $\theta^\pi$ (\texttt{DDPG})
		\State $t \gets t+1$
		\State \textbf{goto} \emph{loop}.
		\EndProcedure
	\end{algorithmic}
\end{algorithm}

Algorithm~\ref{alg:ddpg} gives an outline of the dynamic pricing mechanism in \iris. A new price $p^t$ is selected at each epoch $t$ (\texttt{line 5}) using the policy of the actor $\pi(x | \theta^\pi)$. Some exploration noise $\epsilon$ is added to the price to allow the agent to explore other states. The price is announced to tenants (\texttt{line 7}) and their radio resource requests are collected in return\footnote{An empty request is assumed if a tenant fails to respond at some epoch.}. Based on these requests, \iriss neutral-host allocates the radio resources following the rule in~\eqref{eq:allocated_rbs} (\texttt{line 8}); then calculates the reward $r^t$ and transitions into a new state $x^{t+1}$ (\texttt{line 9}). The parameters of the actor and the critic network are updated based on the DDPG algorithm (\texttt{line 10}), which is the training step, and a new epoch $t+1$ begins. Note that the learning process of \iriss never stops, allowing the pricing mechanism to re-train and adapt to new environments (e.g., as the tenants change their valuations for the radio resources over time). To achieve this \iriss employs a constant learning rate for both the actor and the critic to update policies and, as already mentioned, uses a discounted reward to account for the unpredictability of the environment. 

Regarding the complexity of \iris's dynamic pricing mechanism and based on the description of Algorithm~\ref{alg:ddpg}, it can be seen that \iriss only depends on the number of tenants sharing the infrastructure and is independent of the actual number of UEs associated with each tenant of the system and the traffic that each UE generates. As a result, the computational complexity in the neural networks of the actor and the critic employed by the DDPG algorithm increases linearly in the number of input layer units as the number of tenants grow, with each tenant adding one load $l_i$ and one request $\nu_i$ feature to the input layer. Furthermore, the message exchanges required by the proposed algorithm also increase linearly with the number of tenants. More specifically, assuming $N$ tenants, each round of the algorithm requires $3N$ message exchanges among the neutral-host and the tenants -- $N$ messages for the announcement of the price by the neutral-host to the tenants, $N$ messages for the radio resource requests from the tenants to the neutral-host and $N$ messages for the actual allocation of the radio resources to the tenants by the neutral-host. The computational complexity and communication overhead of \iriss allows it to scale in all settings of practical interest, as demonstrated in the evaluation results presented in Section~\ref{sec:agent_gran}.

\section{Iris Implementation}
\label{sec:impl}

Following the design and dynamic pricing mechanism described above and in order to be able to assess the system's practicality for a real deployment (as explored in \S\ref{sec:agent_gran}), we developed a prototype of \iris, considering LTE as the radio access technology (RAT). To realize RAN slicing, we leveraged the Orion RAN slicing system~\cite{foukas2017orion}, which provides functionally isolated virtual control planes (RAN controllers) for tenants and virtualized radio resources revealed to them through a Hypervisor. Orion is in turn built on top of the OpenAirInterface (OAI) LTE platform~\cite{nikaein2014openairinterface}. OAI has built-in C-RAN support offering three functional splits: lower-PHY, higher-PHY and nFAPI~\cite{nfapi_specs}. Although in principle any of these functional splits could be used in \iris, the Orion implementation is only compatible with the first two. Between them, considering their differences in fronthaul bandwidth requirements (1Gbps with lower-PHY versus 280Mbps for higher-PHY for a 20MHz carrier)~\cite{makris2017experimental,chang2017flexcran}, we opt for the higher-PHY split.  

\noindent\textbf{Edge Cloud Deployment.} To realize the \iriss system design, we leveraged the OpenStack edge cloud deployment of the University of Edinburgh presented in \cite{foukas2018testbed}, which is composed of 5 compute nodes (24-core Xeon CPUs @2.1GHz and 32GB RAM each), optimized for real-time operation (disabled CPU C-states, low-latency Linux kernel, no CPU frequency scaling, VNF CPU pinning). For the RRUs, we employed USRP B210 Software-Defined Radios (SDRs), each interfaced to a small form factor PC (UP board with 4GB of RAM and Intel Atom x5 Z8350 CPUs @1.92GHz), the latter acting as a compute node for running the lower part of the PHY operations and for communication with the BBUs (over Gigabit Ethernet).

\noindent\textbf{Spectrum Manager.} We implemented a prototype Python-based spectrum manager to host the \iriss pricing policy agent, employing an existing implementation of DDPG~\cite{ddpg_impl} that uses Tensorflow for the training of the deep neural networks. Given our hardware constraints, we used a Tensorflow flavor that supports execution only on CPU (no GPU). Regarding the parameters of DDPG, we retained the default values provided in the aforementioned implementation, with the actor and the critic neural networks both having two hidden layers with 400 and 300 units, respectively. For the representation of the state and action space, we employed the OpenAI Gym~\cite{1606.01540} toolkit. \looseness=-1

In a full implementation, shared spectrum support would make use of carrier aggregation. However, given that this functionality is not currently supported by OAI, we used a contiguous band of spectrum to simulate scaling the available shared spectrum up/down (by the spectrum manager through signaling messages to the cell agents).

\noindent\textbf{Cell Agents.} Each cell agent in \iriss is realized via modified Orion Hypervisor in the BBU. The radio resource allocation scheme of \iriss following rule~~\eqref{eq:allocated_rbs} was introduced into the radio resource manager of the Hypervisor. The Hypervisors were interfaced with the spectrum manager using Google Protocol Buffers\footnote{https://developers.google.com/protocol-buffers/} and ZeroMQ\footnote{http://zeromq.org/}. Finally, the protocol used for the communication of the Hypervisors with the virtual control planes of tenants was extended to support the messages required for the shared spectrum price announcements and the radio resource requests of the \iriss tenant agents.\looseness=-1

\noindent\textbf{Tenant Agents.} On the tenant side, we leveraged the Orion virtual control plane implementation, which we extended with the \iriss tenant agents. Note that our design (specifically the dynamic pricing mechanism) is agnostic to tenant behaviors. For the sake of evaluations, our implementation supports a rich set of tenant behaviors enacted in the form of utility functions (described in \S\ref{sec:eval_setup}). However, our implementation could also support other ways of expressing the tenant behaviors.\looseness=-1

\section{Experimental Evaluation}
\label{sec:multi_eval}

\subsection{Evaluation Setup}
\label{sec:eval_setup}

\begin{figure}[t]
	\centering
	\begin{subfigure}[b]{0.4\columnwidth}  
		\centering 
		\includegraphics[width=\textwidth]{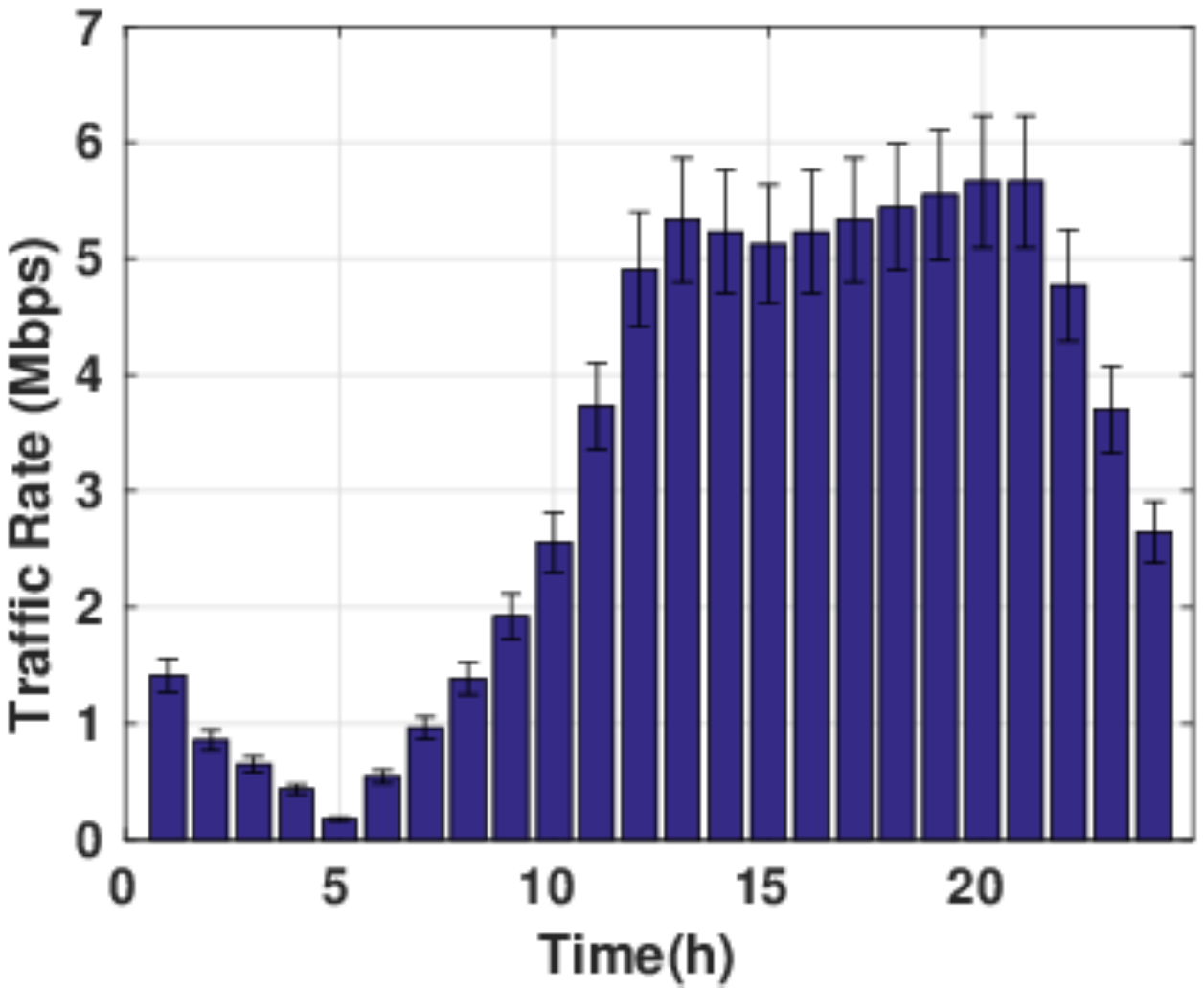}
		\caption{Traffic load}
		\label{fig:traffic_load}
	\end{subfigure}
	~~~~~~~~~~~~
	\begin{subfigure}[b]{0.4\columnwidth}   
		\centering 
		\includegraphics[width=\textwidth]{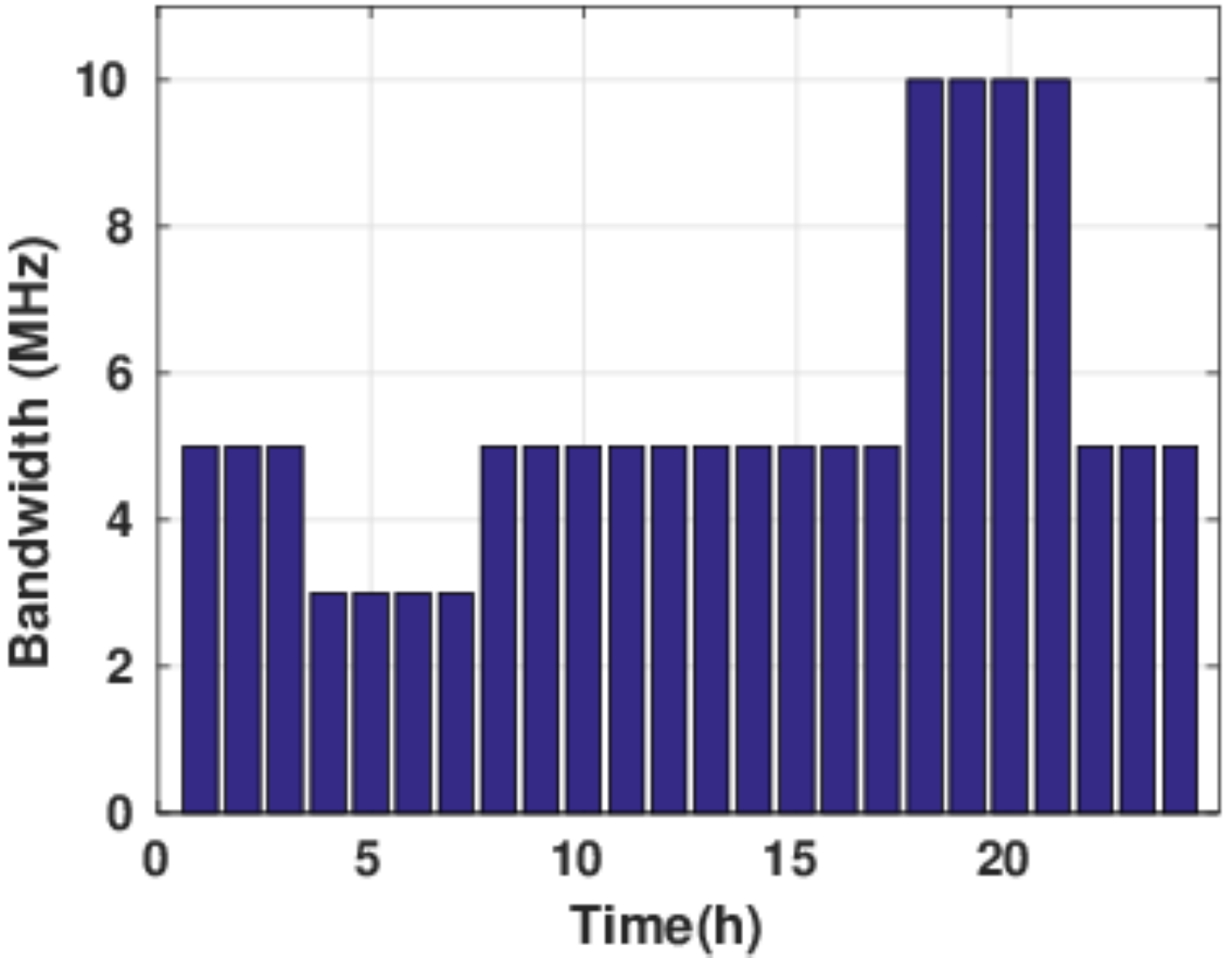}
		\caption{Spectrum availability}
		\label{fig:avail_band}
	\end{subfigure}
	\caption{Daily traffic profile of tenants and spectrum availability profile over a day at the neutral-host.}
	\label{fig:daily_profile}
	\vspace{-3mm}
\end{figure}

For our evaluations, we employ the prototype implementation of \iriss (\S\ref{sec:impl}). The default experimental setup corresponds to 4 tenants per cell. For experiments with a single small cell, real UEs (LTE smartphones and dongles), one per tenant, and the D-ITG traffic generator~\cite{botta2012tool} were used to generate a simulated aggregate UDP traffic for the tenant. UEs were simulated for scenarios with multiple small-cells due to the complexity of managing the experimental setup. It should be noted that even though certain aspects of the system like the generated traffic were simulated (and therefore a simpler simulation setup could also be used), employing the prototype implementation is still crucial for the evaluation, since it provides critical insights regarding the applicability and the overhead of the proposed mechanism in real settings, as discussed in \S\ref{sec:agent_gran}. 

\noindent\textbf{Tenant traffic loads and spectrum availability.}
To model the traffic loads of tenants, we employed the daily aggregate traffic pattern presented in \cite{wang2015understanding} for an entertainment area, a representative indoor environment. We assume that the aggregate incoming hourly traffic of each tenant follows a normal distribution with mean and variance depending on the particular hour in the day considered as shown in Fig.~\ref{fig:traffic_load}. With no real-world data to rely on, we consider a reasonable spectrum availability profile in Fig.~\ref{fig:avail_band}. The idea behind this profile is that the available shared spectrum is re-adjusted (with some delay) by the spectrum manager to approximately match the traffic load. While some of the experiments span the whole day and use the full profiles in Fig.~\ref{fig:daily_profile}, others focus on a particular hour and so use the traffic and spectrum values for that hour. The default evaluation configuration is for the hour starting at 3pm and a cell with 5MHz of available shared spectrum, emulating a CBRS-like service using LTE band 7. Small cells use a SISO transmission mode, which for 5MHz spectrum corresponds to a max throughput of 16Mbps.\looseness=-1 

\noindent\textbf{Dynamic pricing mechanism settings.} The epoch duration is set to 30ms (based on results in \S\ref{sec:agent_gran}), while the presented results correspond to the parameter $\sigma=1$ for \eqref{eq:gaussian} and $\delta=1$ for \eqref{eq:scale_factor}. The price range is set to $\left[0, p_{max}\right]$ with $p_{max}=2500$. In setting the target revenue level $T$, we consider the case where the neutral-host is concerned only with recovering the cost associated with shared spectrum acquisition. Accordingly, we set $T(n) = p_c n,$ where $p_c$ is the cost incurred to the small cell for the acquisition of a single RB. We use the value $p_c=850$, unless explicitly stated otherwise.\looseness=-1

\noindent\textbf{Modeling different tenant behaviors.} As already discussed in Section~\ref{sec:neutral_host_model}, our model is compatible with very general tenant behavior patterns. For the sake of concreteness of our evaluation and to demonstrate the effectiveness of \iris, we modeled the behavior of tenants through dis-utility functions possessing generic parameterizable structure, formulated on the basis of detailed analysis that can be found in the appendix. Specifically, the dis-utility functions have the form \looseness=-1
\begin{equation}
\label{eq:utility}
    \bar{U}(b; d, p) = \Bigl(\alpha (\max(0, d - b))^{\gamma_d} + (pb)^{\gamma_p}\Bigr)^{1/\gamma_p},
\end{equation}
where $p$ denotes the price announced by the pricing policy agent while $d$ and $b$, respectively, represent a tenant's traffic load and requested resources in terms of RBs. All arguments here refer to the same epoch. The parameters $\alpha$, $\gamma_d$ and $\gamma_p$ characterize the individual tenant behavior; the settings of these parameters determine the sensitivity of the dis-utility function to the current load or price and can therefore allow modeling different tenant behaviors and reactions to price changes. These parameters can be modified on-the-fly, allowing the tenants to dynamically change their shared spectrum allocation policy. Raising the sum in~\eqref{eq:utility} to the power~$1/\gamma_p$ expresses the value of the dis-utility in units of ``cost'', bearing the same interpretation for all tenants. This allows introducing the notion of ``total dis-utility'' calculated as the sum of dis-utilities over all tenants. Through \eqref{eq:utility} and given the price $p^t$ and the level of traffic load $l_i^t$, corresponding to $d(l_i^t)$ RBs, the agent of each tenant $i$ requests from the \iriss cell agent the number of RBs that minimizes its dis-utility i.e., $\arg \min_{b} \bar{U_i}(b; d(l_i^t), p)$. 

Based on the above, we created 4 tenant profiles for our evaluations (Table~\ref{tab:tenant_profiles}), using different parameterizations of \eqref{eq:utility} to model different tenant behaviors. The choice of parameters for these profiles was made based on the analysis in the appendix, with the goal of capturing a wide range of sensible and diverse tenant behaviors that would allow a more accurate and realistic evaluation of our mechanism. Unless explicitly stated otherwise, tenants were assigned these profiles in a cyclic manner, i.e., tenant 1 to profile 1, tenant 2 to profile 2, tenant 5 to profile 1 etc. \looseness=-1 

\begin{table}[t]
\scriptsize
{
\centering
\caption{Tenant profiles with different parameterizations of the generic disutility function and the resulting behaviors.}
\label{tab:tenant_profiles}
\begin{tabular}{|c|c|c|c|l|}
\hline
Profile Type & $\alpha$          & $\gamma_p$ & $\gamma_d$ & \multicolumn{1}{c|}{Effect}                                                                                                                                                                                                                               \\ \hline
1. ``Best effort''  & $3.5 \times 10^8$ & 2          & 1          & \begin{tabular}[c]{@{}l@{}}The main focus of the tenant is to maintain a network presence,  providing best effort services \\ with a small amount of radio resources regardless of the price. The tenant is only willing to\\ cover its load for low prices.\end{tabular}                         \\ \hline
2. Price-driven  & $2 \times 10^9$   & 2          & 1          & \begin{tabular}[c]{@{}l@{}}The tenant fully covers its load when the cost is low (e.g. off
peak hours with no congestion). \\For high prices, it only covers part of its load.\end{tabular}                                                                                                                                 \\ \hline
3. Demand-driven   & 0.203             & 1          & 2          & \begin{tabular}[c]{@{}l@{}}The tenant focuses on providing data-demanding services to its users (e.g. video streaming, \\IPTV). In times of high load the tenant is willing to buy a large amount of resources, \\regardless of the price. In other times, the tenant will queue its traffic until the load increases\\ enough to buy in bulk. \end{tabular} \\ \hline
\begin{tabular}[c]{@{}c@{}}4. ``Medium'' QoS \\ level \end{tabular} & $1.1 \times 10^5$ & 2          & 2          & \begin{tabular}[c]{@{}l@{}}Tenant tries to provide a medium level of service, asking a price-dependent fraction\\ of its load.\end{tabular}                                                                           \\ \hline
\end{tabular}
}
\end{table}

\subsection{Deep Learning Benefits, Feasibility and Scalability}
\label{sec:agent_gran}

\begin{figure}[t]
	\centering
    \begin{subfigure}[b]{0.45\columnwidth}   
		\centering 
		\includegraphics[width=\textwidth]{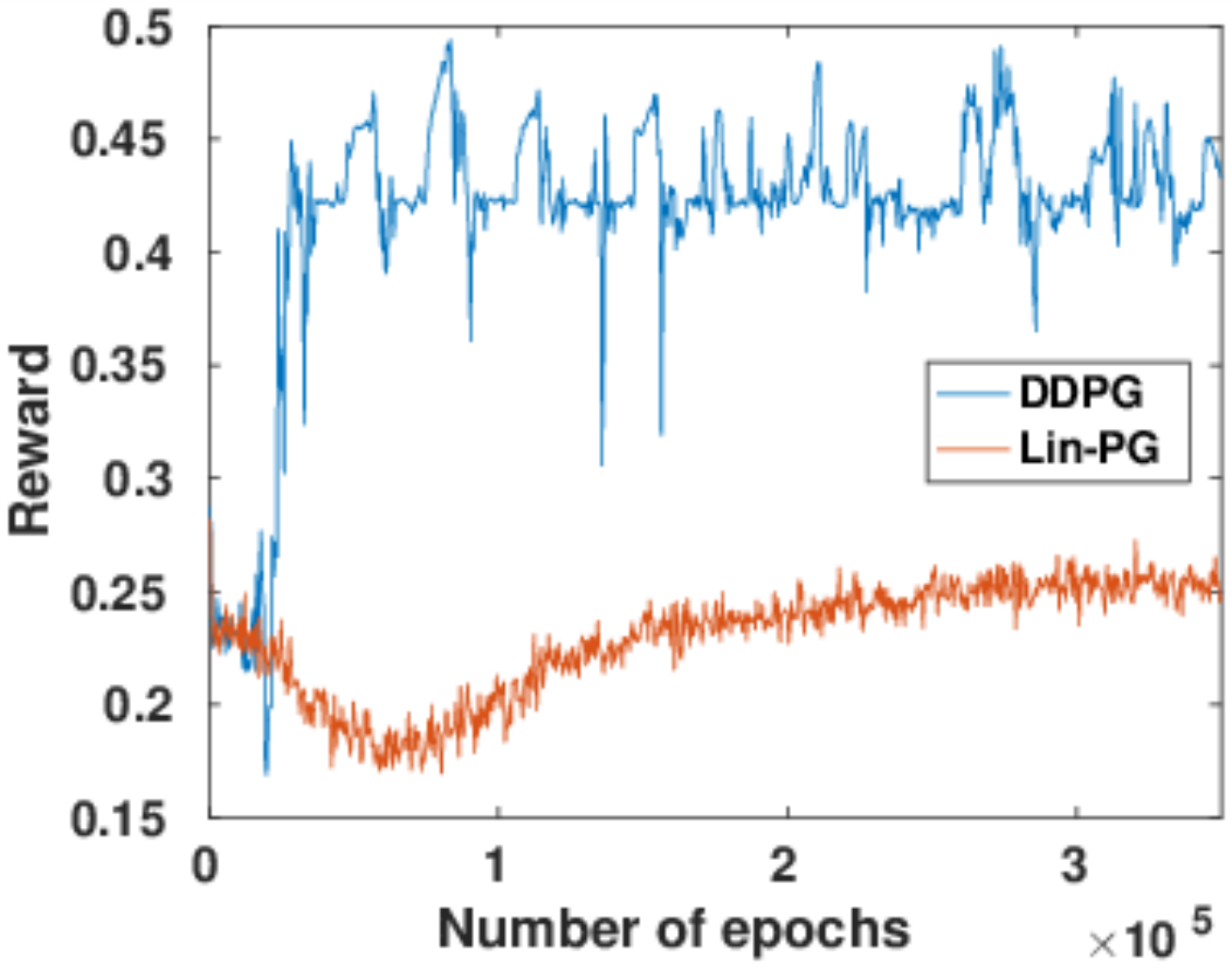}
		\caption{Comparison of DDPG and policy gradient algorithm using linear function approximators.}
		\label{fig:algo_comparison}
	\end{subfigure}
	~~~~~~~~
	\begin{subfigure}[b]{0.45\columnwidth}  
		\centering 
		\includegraphics[width=\textwidth]{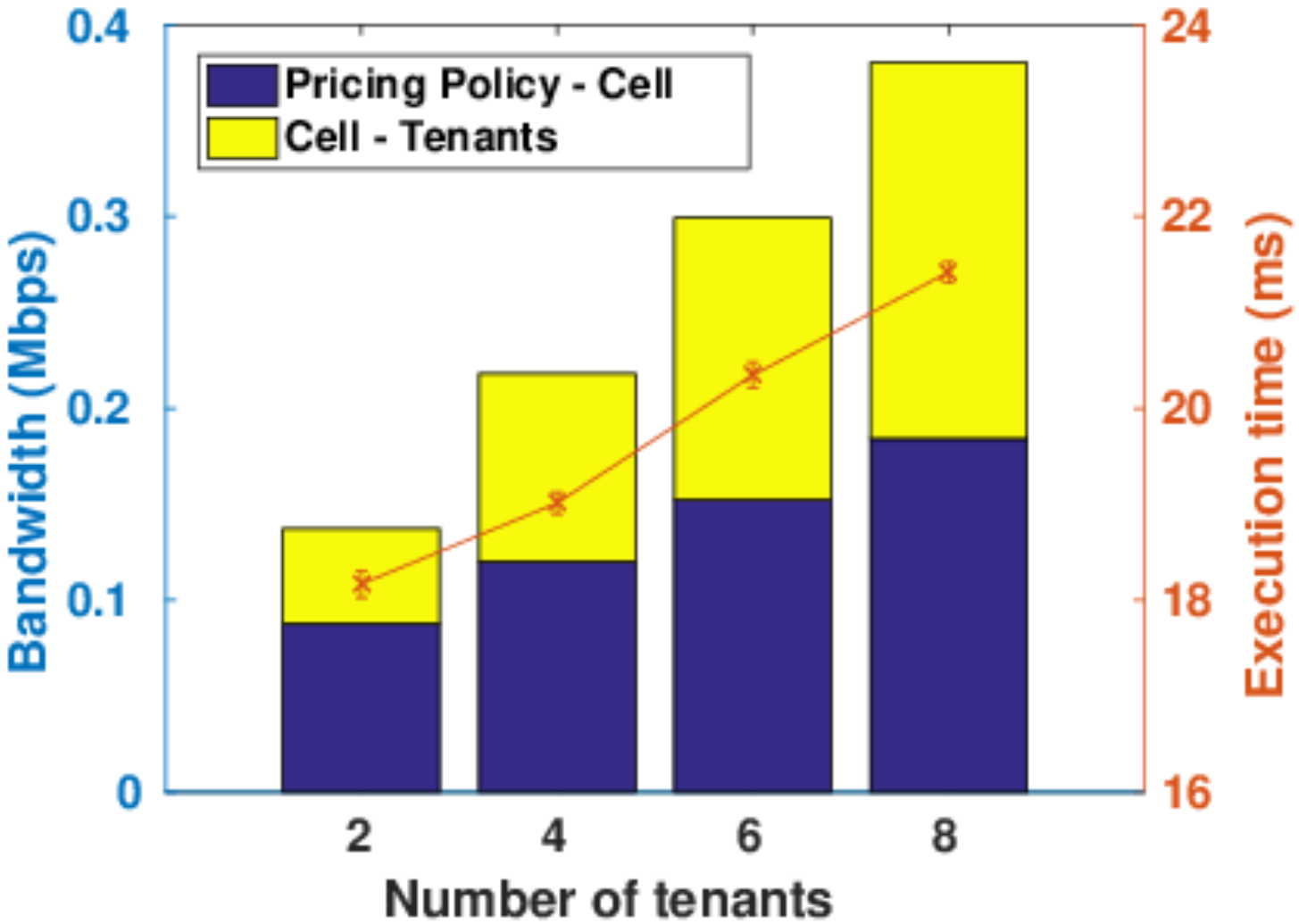}
		\caption{\iriss single training step time and bandwidth requirements for \iriss agents message exchanges}
		\label{fig:exec_time}
	\end{subfigure}
	\caption{Benefits/feasibility of deep reinforcement learning.}
	\label{fig:ddpg_applicability}
	\vspace{-2mm}
\end{figure}

We begin by examining the choice of employing deep reinforcement learning against simpler reinforcement learning algorithms for solving the problem formulated in \S\ref{sec:form_multi}. We compare the performance of \iriss when using DDPG against the stochastic policy gradient algorithm of~\cite{degris2012model}, which employs linear function approximators for the actor and the critic (Lin-PG). As illustrated in Fig.~\ref{fig:algo_comparison}, DDPG converges faster than Lin-PG and attains a much higher overall reward, both of which are critical characteristics for the success of the proposed mechanism in a real deployment.\looseness=-1

Another very important aspect in terms of the mechanism's practicality is that the benefits of deep reinforcement learning and the requirement for the real-time communication of tenants with the neutral-host should not come at the expense of the system's feasibility and scalability. For this reason, we use our prototype implementation to evaluate the performance of \iriss for deployments supporting up to 8 tenants, in order to capture the effects of the scale that we envision for most practical deployments of the system. Our results are in accordance with the analysis of the computational complexity and communication overhead presented earlier in Section~\ref{sec:ddpg}. As illustrated in Fig.~\ref{fig:exec_time}, the time required to calculate the new parameters of the actor and the critic functions by DDPG in a single training step increases linearly with the number of tenants, but remains below 22ms even for 8 tenants. This linear effect correlates with the computational complexity introduced by the linear increase in the number of input layer units in the neural networks of the actor and critic as the number of tenants grow and is in line with the complexity discussion in Section~\ref{sec:ddpg}. When setting the epoch duration, the additional overhead introduced by the message exchanges between the various \iriss agents should also be taken into account. Therefore, setting the epoch duration to 30ms is a reasonable choice (also used in all our consecutive experiments), that provides a very fine granularity in terms of the neutral-host agent's training speed. Offloading the training computations to GPUs (rather than using the CPU as in our current implementation) can potentially lead to significant reductions in the execution time, which in turn allows a lower epoch duration and enables \iriss to learn even faster. The bandwidth requirements of \iriss for the message exchanges between pricing policy and cell agents as well as between cell and tenant agents are also illustrated in Fig.~\ref{fig:exec_time}. These requirements are minimal (less than 0.4Mbps) for all practical deployment scenarios of up to 8 tenants and posing a negligible overhead to the edge cloud deployment.

It should be noted that, as discussed in Section~\ref{sec:ddpg}, the results illustrated in Fig.~\ref{fig:exec_time} only depend on the number of tenants sharing the infrastructure and are independent of the amount of traffic and the way it was generated (simulated or real traffic) or of the actual number of UEs associated with the tenants of the system. Therefore, our prototype implementation provides us with a very accurate depiction of \iris's overhead, demonstrating the feasibility of the system's deployment. 

\subsection{Characterizing Iris Spectrum Management}
\label{sec:iris_eval_behavior}

Next, we characterize the behavior of the proposed allocation mechanism, considering various aspects of the dynamic pricing model and their effects to reinforcement learning in terms of efficiency and time convergence.  \newline

\noindent \textbf{Learning behavior for various traffic loads.} We evaluate the learning behavior of the pricing policy agent for four tenants under three different scenarios, each considering a cell with a different aggregate traffic load: (i) a congested cell (Cell 1), corresponding to the conditions at 3pm from the daily traffic and spectrum availability profiles of Fig.~\ref{fig:daily_profile}; (ii) an uncongested cell (Cell 2) with low traffic load, corresponding to 8am; and (iii) a cell with high traffic loads (Cell 3) but not in congested state, corresponding to 11am.

\begin{figure}[t]
	\vspace{-2.5mm}
	\centering
	\begin{subfigure}[b]{0.45\columnwidth}  
		\centering 
		\includegraphics[width=\textwidth]{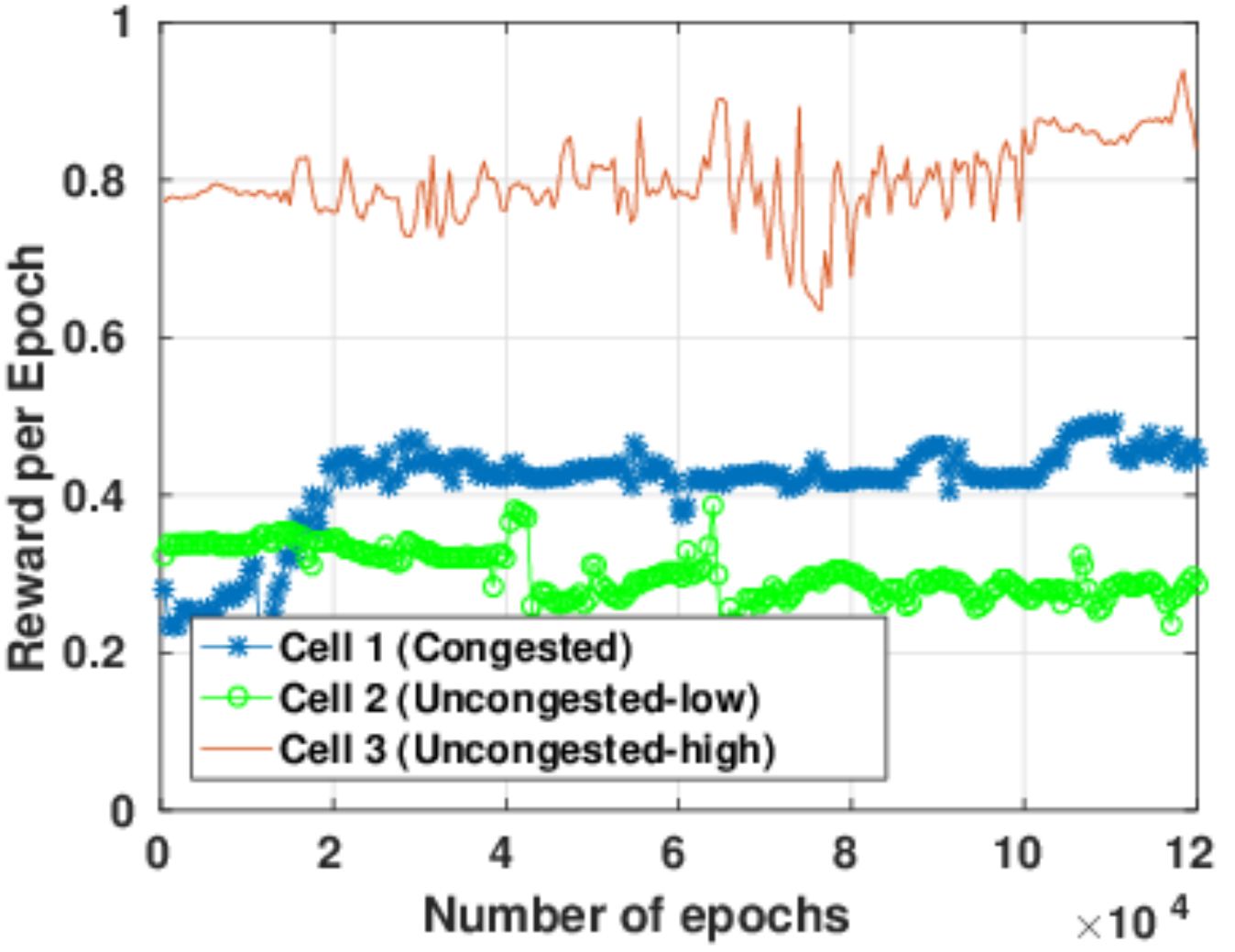}
		\caption{Reward received by neutral-host per pricing epoch.}
		\label{fig:convergence_time}
	\end{subfigure}
	~~~~~~~~
	\begin{subfigure}[b]{0.45\columnwidth}   
		\centering 
		\includegraphics[width=\textwidth]{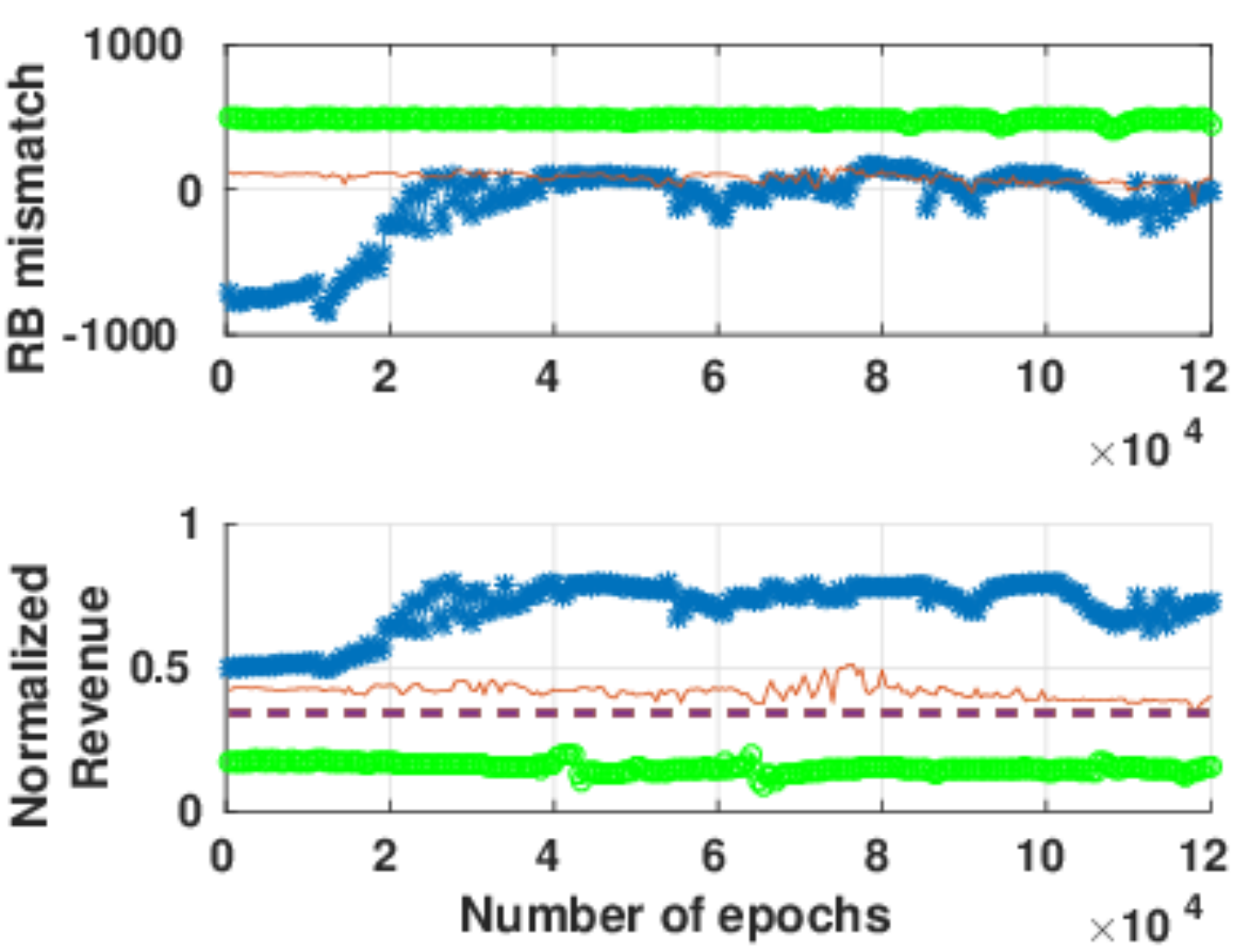}
		\caption{RB mismatch and achieved revenue vs target revenue (dashed line).}
		\label{fig:convergence_mismatch}
	\end{subfigure}
	\caption{Learning behavior of pricing policy agent for cells with different levels of congestion/loads.}
	\label{fig:convergence_analysis}
	\vspace{-2mm}
\end{figure}

Fig.~\ref{fig:convergence_analysis} shows the results for the reward obtained by the pricing policy agent for each of the three scenarios considered. It also shows the average mismatch between the amount~$n^t$ of RBs available during an epoch and the amount $\sum_{i \in I}\nu_i^t$ collectively requested by all tenants, as well as the actual (target) revenue received (set) by the neutral-host, normalized by the maximum possible revenue for an epoch (equal to $p_{max} n^t$). In view of this normalization, the value of the target revenue $T$ (dashed line) maintains the same value (equal to $p_c/p_{max}$) for all three scenarios.\looseness=-1

In the congested (cell 1) case, the agent begins with a very high RB mismatch, which gets close to 0 after the first 20000 epochs, indicating that the pricing policy agent is effectively and quickly learning how to control the requests of the tenants. The neutral-host achieves this by increasing the price of the RBs as evident from its revenue increase. It should also be noted that the big difference between achieved and target revenue levels has an effect to the overall reward of the neutral-host (through function~$g$), which converges to a value that is less than half of the max reward 1.\looseness=-1

For cell 2, the mismatch is always positive (underutilization) and close to 500, since the load of the tenants is very low and the demand can never match the supply. Due to the very low load, it is infeasible for the neutral-host to fully recover its costs for acquiring the shared spectrum, regardless of the pricing policy it follows, something reflected in its reward that is scaled down by \eqref{eq:scale_factor}.\looseness=-1 

Finally, in the case of cell 3 the agent presents a stable behavior, with its RB mismatch and revenue remaining relatively static throughout the experiment. The aggregate traffic of the tenants requires an amount of RBs that is roughly equal to the RBs that are available in the system, while the revenue that is achieved by the neutral-host is very close to the target revenue, leading to an overall reward for the neutral-host agent that is much closer to the max compared to the other two cases. 

\begin{figure}[t]
	\centering
	\begin{subfigure}[b]{0.45\columnwidth}  
		\centering 
		\includegraphics[width=\textwidth]{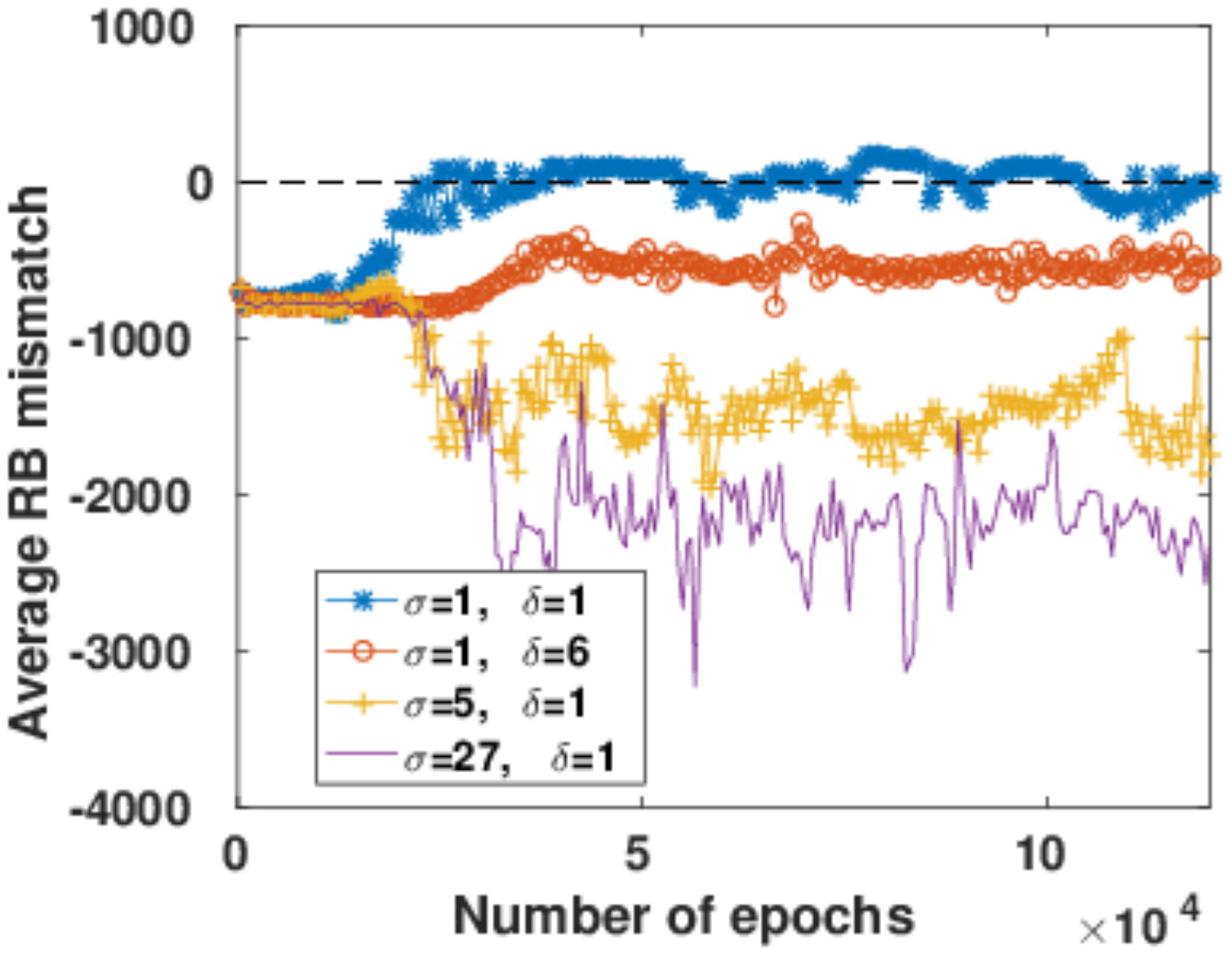}
		\caption{RB mismatch.}
		\label{fig:offset_reward_comparison}
	\end{subfigure}
	~~~~~~~~~
	\begin{subfigure}[b]{0.45\columnwidth}   
		\centering 
		\includegraphics[width=\textwidth]{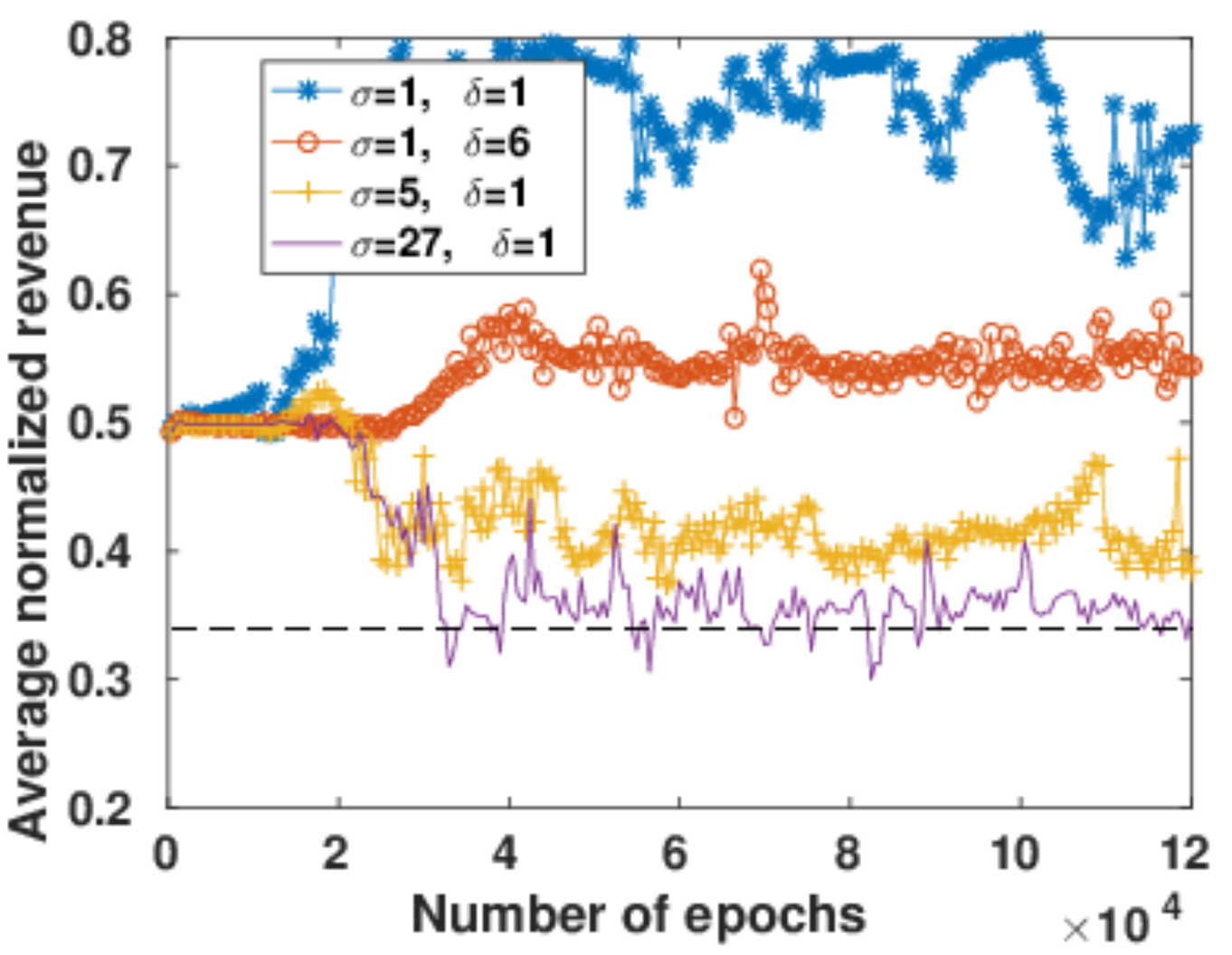}
		\caption{Neutral-host revenue.}
		\label{fig:rev_reward_comparison}
	\end{subfigure}
	\caption{Effect of reward function parameters on pricing behavior.}
	\label{fig:reward_comparison}
	\vspace{-2mm}
\end{figure}

\noindent \textbf{Effect of reward function.} We explore how
the configuration of the reward function affects the neutral-host's
learning process, considering four variants of the reward function with different combinations for the parameters $\sigma$ and $\delta$, as shown in Fig.~\ref{fig:reward_comparison}. In accordance with the discussion in \S\ref{sec:form_multi}, we can observe that as the value of $\sigma$ increases, the RB mismatch becomes more unstable and/or higher from one round to the next (Fig.~\ref{fig:offset_reward_comparison}), but at the same time the achieved revenue gets closer to the target revenue $T$ (dashed line in Fig.~\ref{fig:rev_reward_comparison}). The reason for this behavior is that higher values of $\sigma$ can tolerate higher RB mismatches (since the bell curve of \eqref{eq:gaussian} widens). Therefore, even high mismatches yield relatively significant reward contributions from~\eqref{eq:gaussian}, something that simplifies the pricing decision, by making the pricing policy to be mainly driven by the other factor \eqref{eq:scale_factor} of the reward function. 
The results in Fig.~\ref{fig:reward_comparison} also indicate (as per discussion in \S\ref{sec:form_multi}) that increasing the value of $\delta$ also promotes a closer match between actual and target revenue levels (driven by a more significant effect of \eqref{eq:scale_factor}). Naturally, this has also an effect in the achieved RB mismatch, which increases. This is because the effort to match actual and target revenue levels triggers a lower price per RB, subsequently leading to an increased demand by the tenants.

\begin{figure*}[ht]
	\centering
	\begin{subfigure}[b]{0.45\columnwidth}  
		\centering 
		\includegraphics[width=\textwidth]{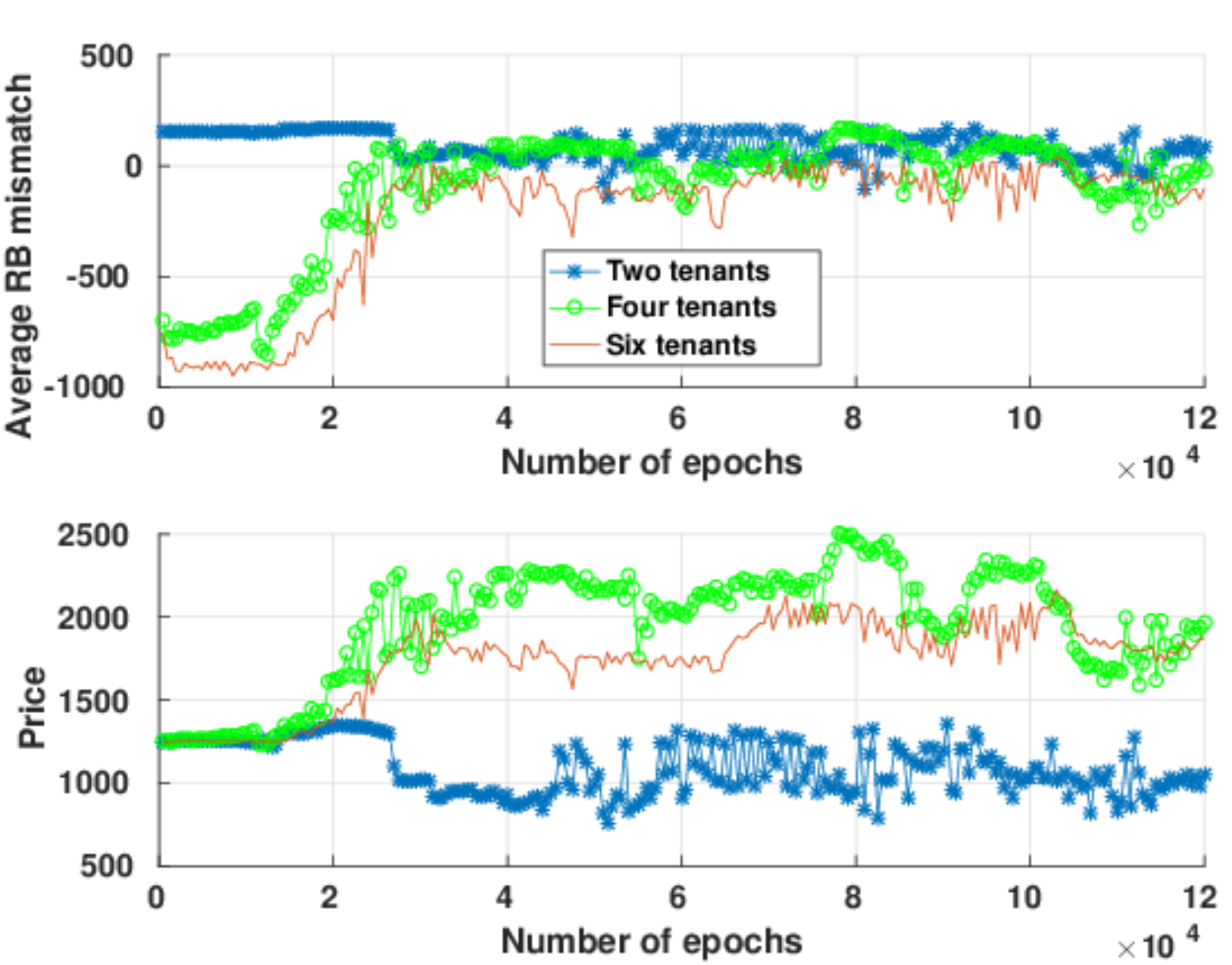}
		\caption{Effect of number of tenants in learning.}
		\label{fig:convergence_num_tenants}
	\end{subfigure}
		~~~~~~~~~
	\begin{subfigure}[b]{0.45\columnwidth}   
		\centering 
		\includegraphics[width=\textwidth]{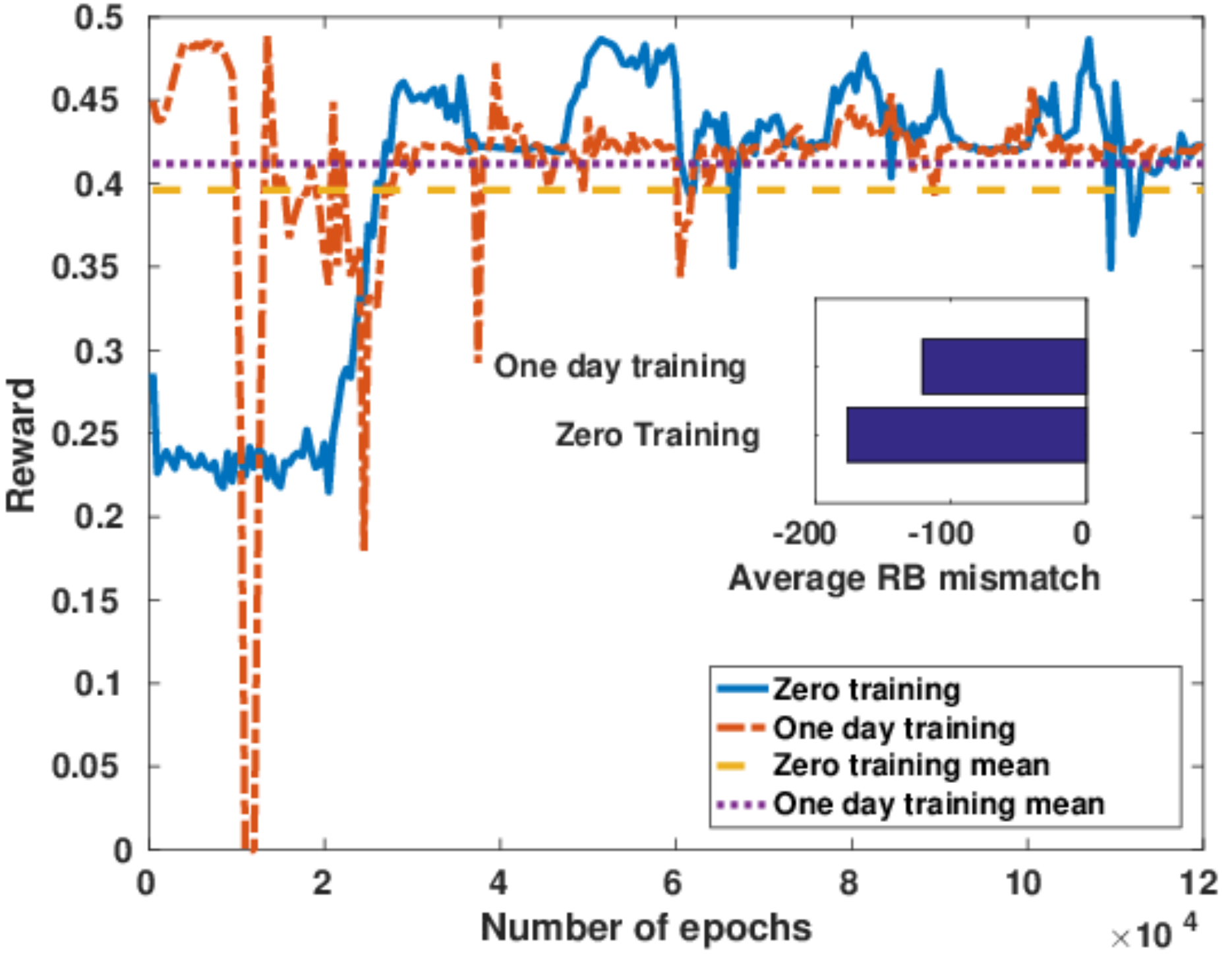}
		\caption{Prior training effect to pricing mechanism.}
		\label{fig:effect_prior_training}
	\end{subfigure}
	\begin{subfigure}[b]{0.45\columnwidth}   
		\centering 
		\includegraphics[width=\textwidth]{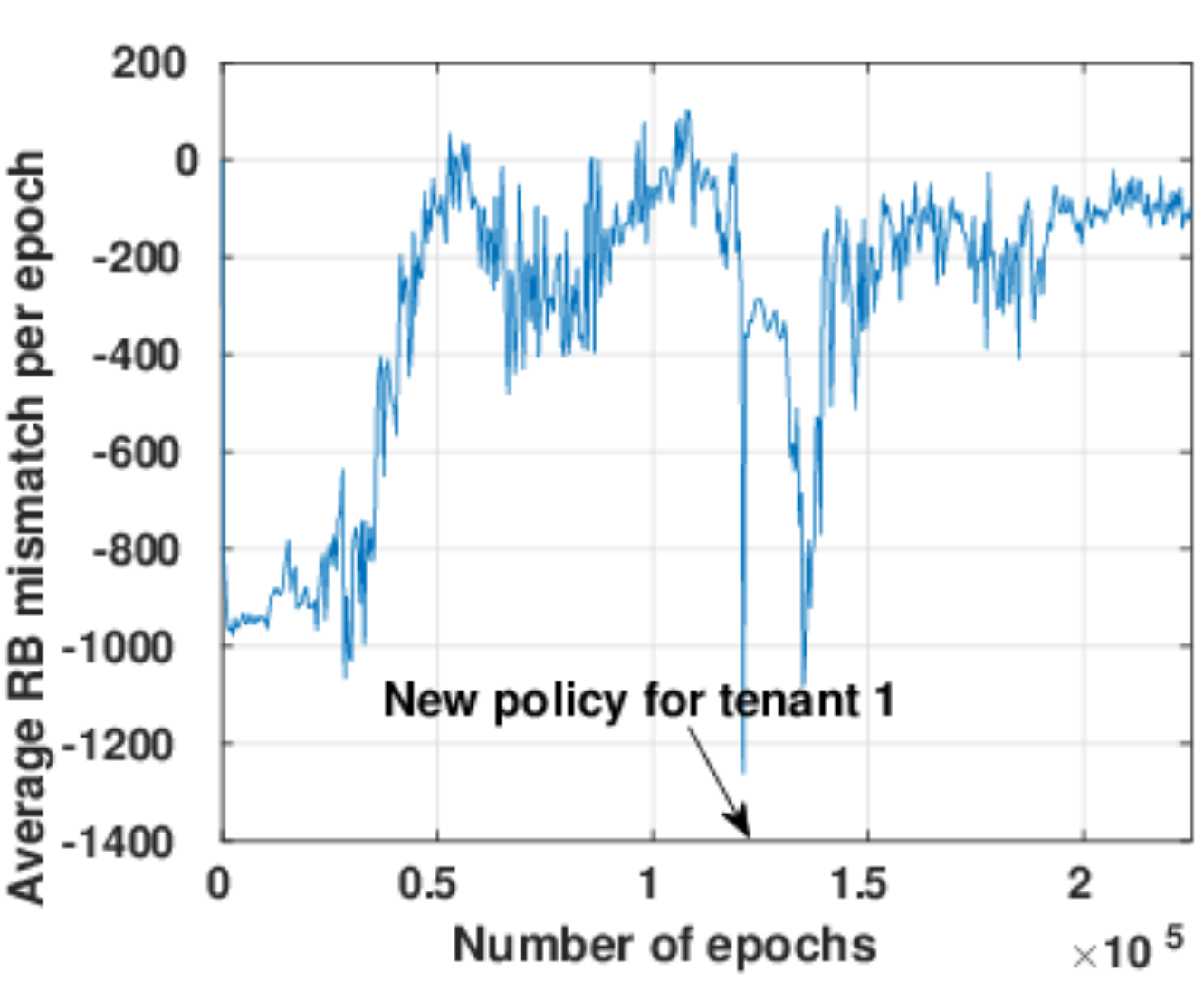}
		\caption{Adaptation to dynamic policy changes.}
		\label{fig:convergence_policy_change}
	\end{subfigure}
    	~~~~~~~~~
    \begin{subfigure}[b]{0.45\columnwidth}   
		\centering 
		\includegraphics[width=\textwidth]{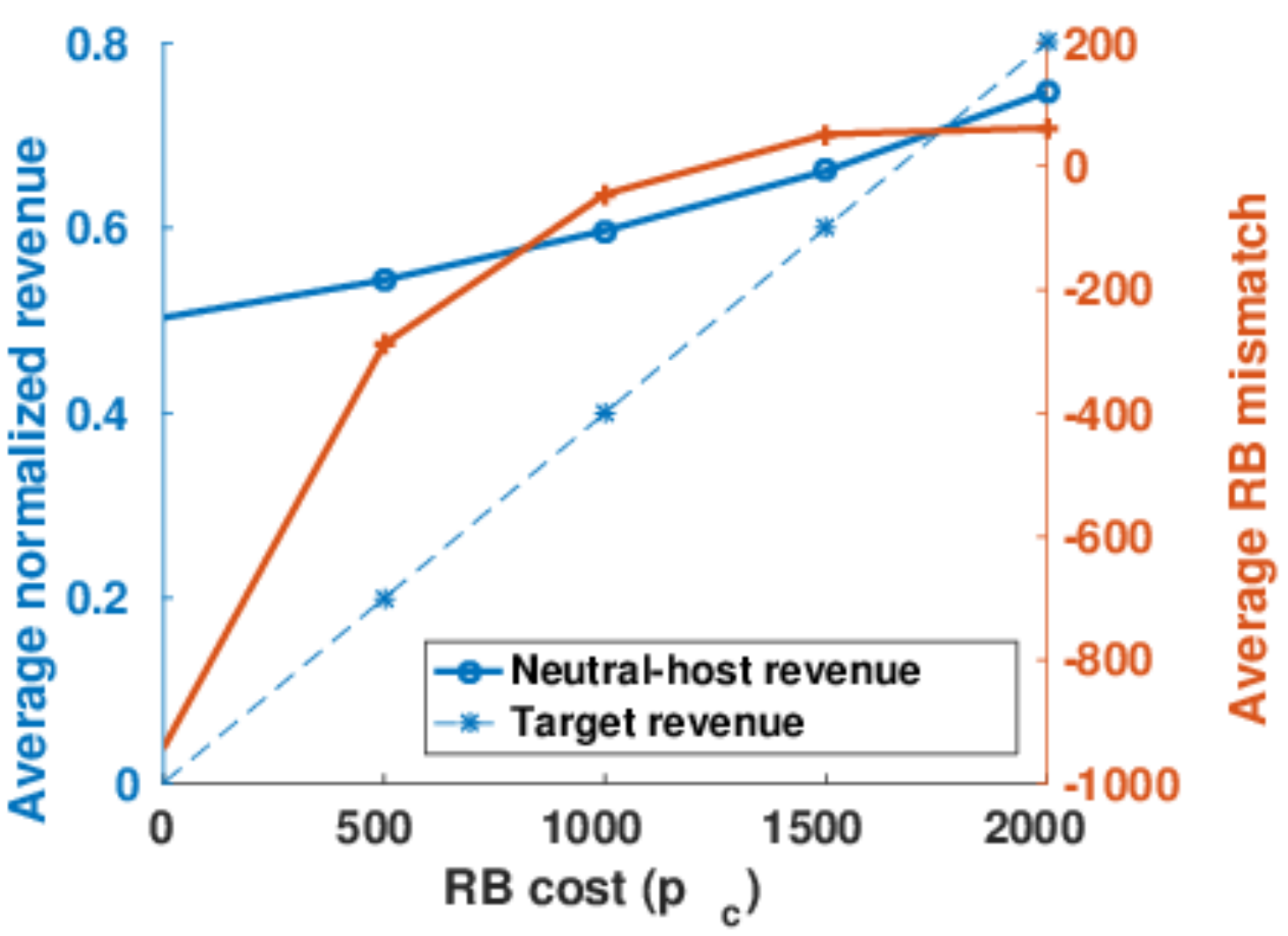}
		\caption{Impact of target revenue.}
		\label{fig:price_effect}
	\end{subfigure}
	\caption{Behavior of \iriss pricing mechanism under different conditions: varying number of tenants; prior training; dynamic tenant policy changes; and different costs for acquiring shared spectrum.}
	\label{fig:var_comparison}
	\vspace{-2mm}
\end{figure*}

\noindent\textbf{Different number of tenants.} We evaluate the learning behavior of the dynamic pricing mechanism as the number and behavior of active tenants vary (recall that the behavior of each tenant depends on its index, as explained in \S\ref{sec:eval_setup}). We consider three cases with two, four and six tenants and a congested cell with the aggregate traffic of 16Mbps across all tenants with equal levels of traffic. Fig.~\ref{fig:convergence_num_tenants} shows the average RB mismatch and the pricing choices of the neutral-host. For four and six tenants the system begins with a negative mismatch, while for two tenants with a positive mismatch. This is related to the effect of neutral-host's pricing choices to the tenants, given their loads and shared spectrum allocation profiles. For two tenants, the initial prices are considered high, leading to the underutilization of the resources, despite the cell congestion. On the other hand, for four and six tenants and given the increased competition, the price is low, leading to excess demand. In all cases, the agent adapts and discovers an appropriate pricing policy to minimize the mismatch.\looseness=-1

\noindent\textbf{Effect of prior training.} We evaluate the effect of prior training to the results achieved by the neutral-host. We perform an experiment for two consecutive days, starting from a state of no training and focus on the results obtained during the same hour of the day (3pm from the daily profile). As it can be seen in Fig.~\ref{fig:effect_prior_training}, the second day yields improved results compared to the first day (zero training), which is evident both from the better mean reward (the brief drops of the instantaneous reward, lasting for less than a minute each, are inconsequential), as well as from the reduced average RB mismatch during the second day. However, the differences between the two days are small, demonstrating the effectiveness of the mechanism even without any substantial prior training.   

\noindent\textbf{Effect of dynamic policy changes.} This experiment demonstrates the adaptiveness of the dynamic pricing mechanism when tenants make policy changes dynamically. In this scenario, the experiment runs for 120000 epochs using the default tenant profiles. When this period elapses, and once the agent has identified an appropriate pricing policy for the given load, the first tenant's policy changes to profile \#2 (Table~\ref{tab:tenant_profiles}). This leads to a temporary failure of the agent to appropriately price the available radio resources, which is translated into a major RB mismatch (Fig.~\ref{fig:convergence_policy_change}). However, after about 30000 epochs (150000 epochs in the experiment), the neutral-host agent manages to re-adapt to the new behavior of tenant 1.  

\noindent\textbf{Effect of target revenue level/\discretionary{}{}{}shared spectrum acquisition cost.} 
We evaluate the effect of the target revenue level to the pricing decisions. Since we employ target revenue levels of the form $T(n)=p_c n^t$, changes to the target revenue level correspond to changes in the shared spectrum acquisition price $p_c$. We proceed by applying different values of~$p_c$ and comparing the actual (target) revenues received (set) by the neutral-host, again both normalized by the maximum possible revenue for an epoch. As it can be seen in Fig.~\ref{fig:price_effect}, both revenue levels increase along with~$p_c$. For the lower acquisition prices, and due to the high load (traffic at 3pm) and the network congestion, tenants are willing to buy the RBs in prices much higher than $p_c$. Therefore, the neutral-host finds a balance on the goals of \eqref{eq:neutral_host_reward} by announcing lower prices (creating excess demand) in order to bring the actual and target revenue levels as close as possible (avoiding the overcharging of tenants for the resources). On the other hand, as the acquisition price increases, the gap between the target and actual revenue levels decreases, with the target surpassing the actual revenue for $p_c=2000$. At the same time, the RB mismatch becomes smaller and turns from excess demand to excess supply (positive mismatch) for $p_c=1500$ and $p_c=2000$. This is because the neutral-host, driven by its reward function, learns pricing policies that make the tenants buy less RBs on average, but at higher prices.\looseness=-1

\subsection{Comparison with Alternative Approaches}
\label{sec:eval_alt}

We compare the performance of the \iriss dynamic pricing mechanism with alternative approaches in terms of the benefits provided to tenants. We consider the traffic generated for a whole day (full profile presented in Fig.~\ref{fig:daily_profile}) for four tenants with their behaviors defined in Table~\ref{tab:tenant_profiles}. The agent is evaluated against other schemes without any prior training. This worst-case scenario is important to benchmark the effectiveness of the neutral-host agent's operation in volatile environments.

We compare the dynamic pricing mechanism of \iriss against the distributed optimization algorithm proposed in \cite{low1999optimization}. Based on that, the neutral-host iteratively adjusts the price of the available RBs in order to control the behavior of tenants that are driven by the goal of minimizing their dis-utilities. Assuming a static environment, the algorithm in \cite{low1999optimization} has been shown to converge to an optimal solution in terms of the utilization of the available resources, but does not inherently capture the requirement of \iriss regarding the revenue target of the neutral-host. For this, we consider two variants of \cite{low1999optimization}: (i)~the vanilla version in which the neutral-host does not set a reserve price for the resources it distributes to the tenants (Distributed No Reserve
Price---DNRP) and; (ii)~a modified version, in which the neutral-host sets a reserve price equal to the cost of a resource block ($p_c=850$), in an attempt to avoid experiencing losses (Distributed Reserve Price---DRP).\looseness=-1

Another alternative we compare against and which could be viewed as a variant of the optimal solution is an unrealistic myopic pricing scheme in which the neutral-host knows the dis-utility functions of the tenants. Using this knowledge, it determines at each epoch, the price to charge the tenants by minimizing the sum of tenant dis-utilities, subject to the resource availability constraint and the requirement that the neutral-host matching or exceeding the target revenue level, i.e.,\looseness=-1
\begin{gather*}
  \min_{p, \vec{\nu}}\sum_{i \in I}\bar U_i (\nu_i; d, p) \\ 
  \text{s.t.}\quad
p\sum\nu_i \geq T(n), \quad \sum\nu_i \leq n, \quad \nu_i \geq 0, \quad \forall i \in I
\end{gather*}
The neutral-host allocates the resources myopically during each epoch (in the sense that it views each epoch in isolation) so that it does not incur losses even in the short-term. A side-effect of this is that under very low traffic load, the neutral-host forces the tenants to buy more resources than they actually need, to recover the acquisition cost for the spectrum. 
It is noted that the comparative evaluation does not consider as a baseline the unrealistic but ``optimal'' solution which optimizes the allocation of resources considering the network dynamics (traffic, spectrum cost, tenant behaviors) throughout the whole day. The complex modeling requirements accounting for the dependency across epochs, the high computational complexity of obtaining the optimal solution and the fact that this needs to be performed over and over again in time makes it impractical even as a baseline.

In addition to the myopic scheme outlined above, we also consider four static pricing schemes, where the price announced by the agent during each epoch $t$ is fixed to $p_{max}/8 \approx 312$ (Static Low), $3p_{max}/8 \approx 937$ (Static Med-L), $5p_{max}/8 \approx 1562$ (Static Med-H) and $7p_{max}/8 \approx 2187$ (Static High) correspondingly, to capture the whole range of possible prices.\looseness=-1

\begin{figure}[t]
	\centering
	\includegraphics[width=0.7\columnwidth]{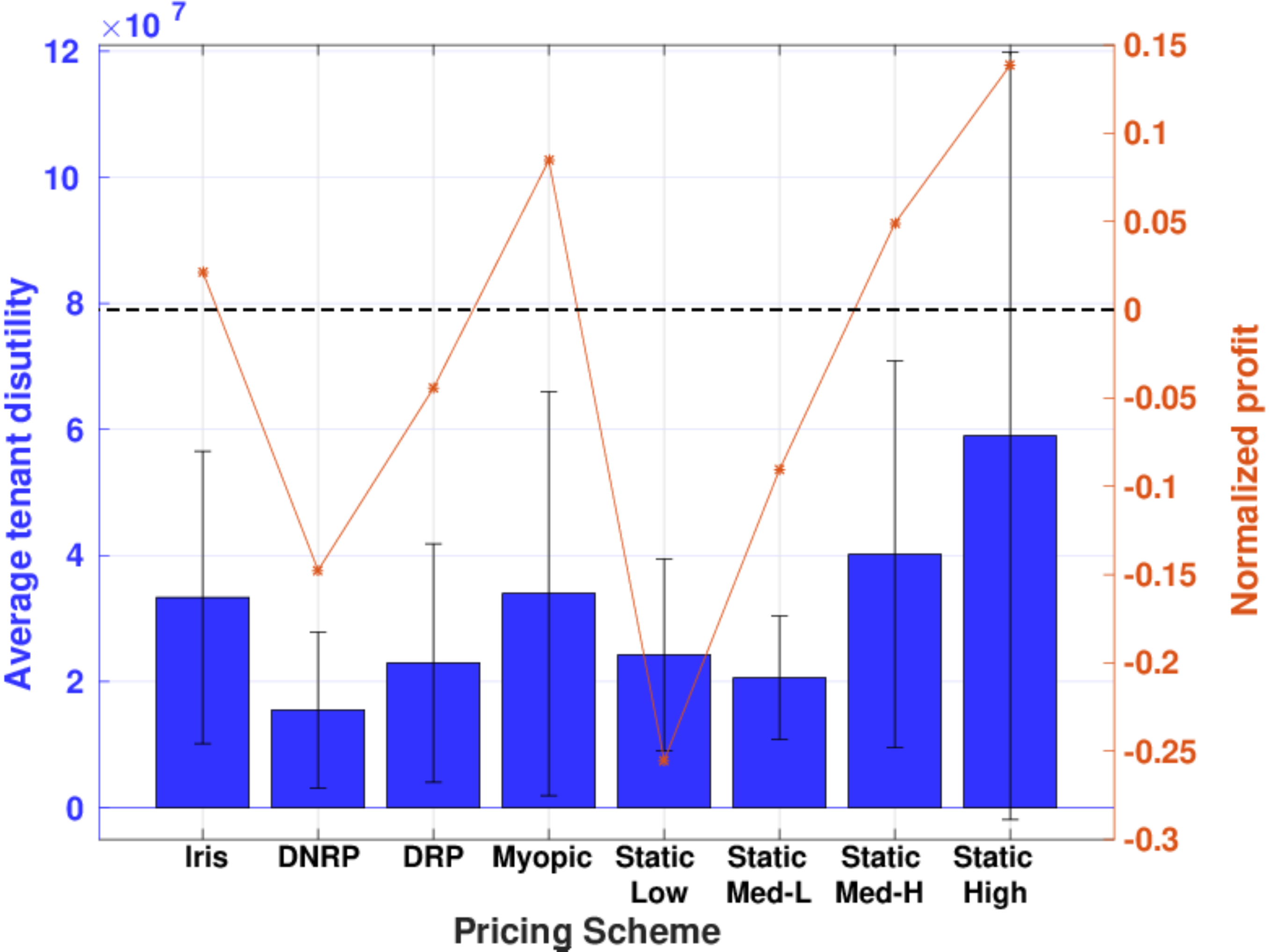}
	\caption{Comparison of \iriss with alternative approaches in terms of total dis-utility of tenants and profit of neutral-host.}
	\label{fig:overall_disutil}
	\vspace{-2mm}
\end{figure}

We begin by looking at the average dis-utility of the tenants for each pricing scheme and the total normalized revenue above the target level made by the neutral-host (Fig.~\ref{fig:overall_disutil}). The revenue 
is normalized by the maximum possible revenue of the neutral-host, i.e., selling all the available resources at the max price of $p_{max}$. In terms of the dis-utility, we can observe that \iriss performs worse than DNRP and DRP as well as two of the lowest static pricing schemes (Static Low and Static Med-L). However, through the revenue results, we observe that for those four schemes the neutral-host experiences losses (negative profit), disincentivizing the neutral-host to provide its service in the first place. This could have been avoided if the pricing policy dynamically adapted not only based on the utilization of the resources, but also based on the revenue target set by the neutral-host.\looseness=-1

\begin{figure*}[t]
	\centering
		\includegraphics[width=0.9\textwidth]{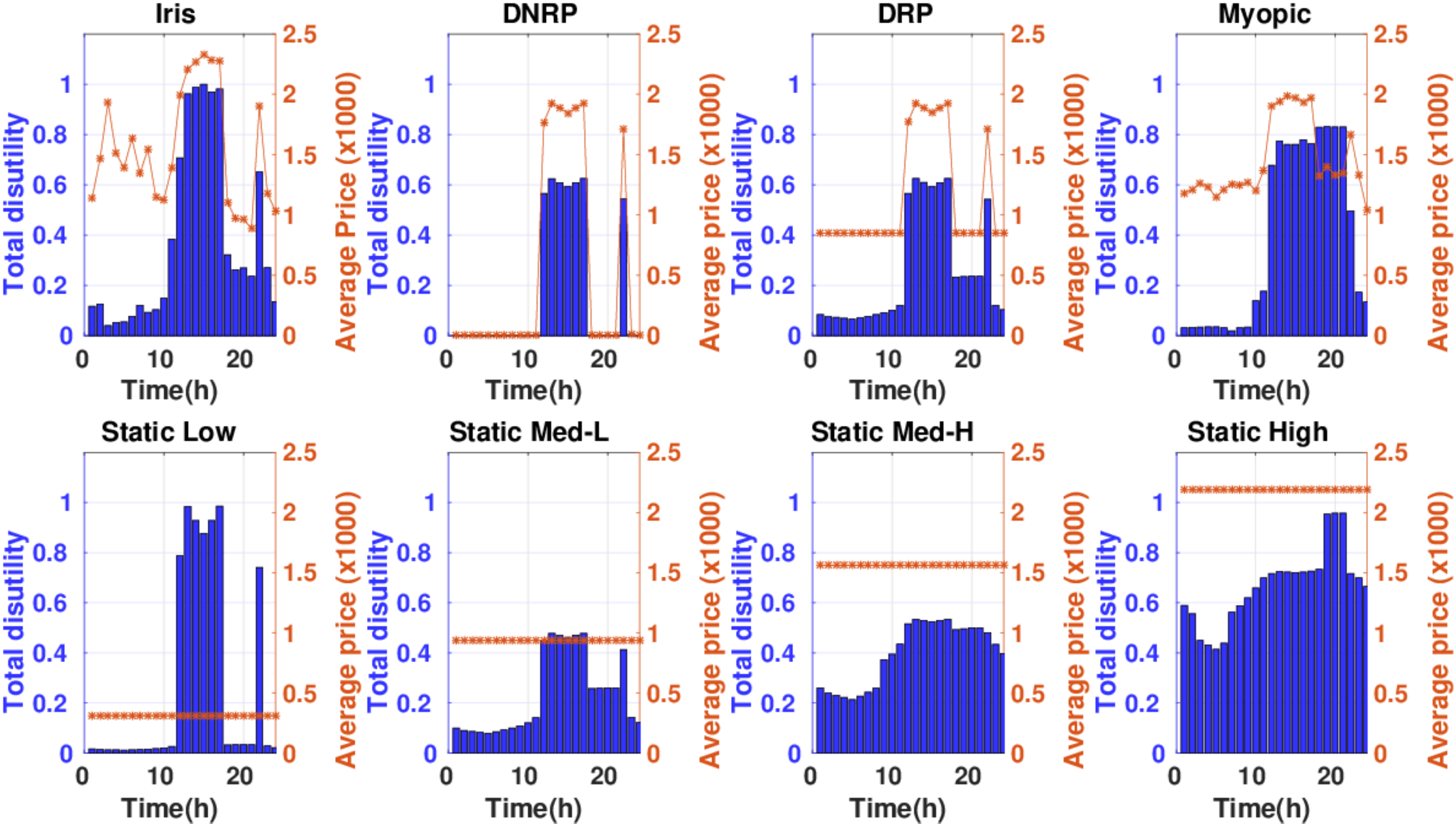}
		\caption{Hourly breakdown of tenants' total dis-utility and average price selected by neutral-host in \iriss against alternative approaches.}
		\label{fig:analytic_day_disutil}
		\vspace{-2mm}
\end{figure*}

The results are opposite for the myopic and the higher static pricing schemes (Med-H and High) in that with these schemes the neutral-host obtains a revenue that is higher than the set target at the expense of a higher tenant dis-utility compared to \iris. For the static pricing schemes, this is due to the inability of the pricing mechanism to adapt to the traffic loads, charging high prices even at times of no congestion (e.g., 1am-10am when the traffic load is low or 6pm-9pm when there is abundance of spectrum). For the myopic scheme, however, this is due to the neutral-host agent forcing tenants to buy resources not needed to myopically recover its spectrum acquisition cost within each epoch. These behaviors are better seen in the hourly breakdown of the tenants' dis-utilities and corresponding prices decided by the neutral-host as shown in Fig.~\ref{fig:analytic_day_disutil}. The \iriss dynamic pricing mechanism manages to draw a balance between the needs of the tenants and the neutral-host more effectively, learning the right pricing policy that keeps the tenants as satisfied as possible, but without incurring a low revenue that would disincentivize the neutral-host. \looseness=-1

\begin{figure}[t]
\centering
	\begin{subfigure}[b]{0.45\columnwidth}   
		\centering 
		\includegraphics[width=\textwidth]{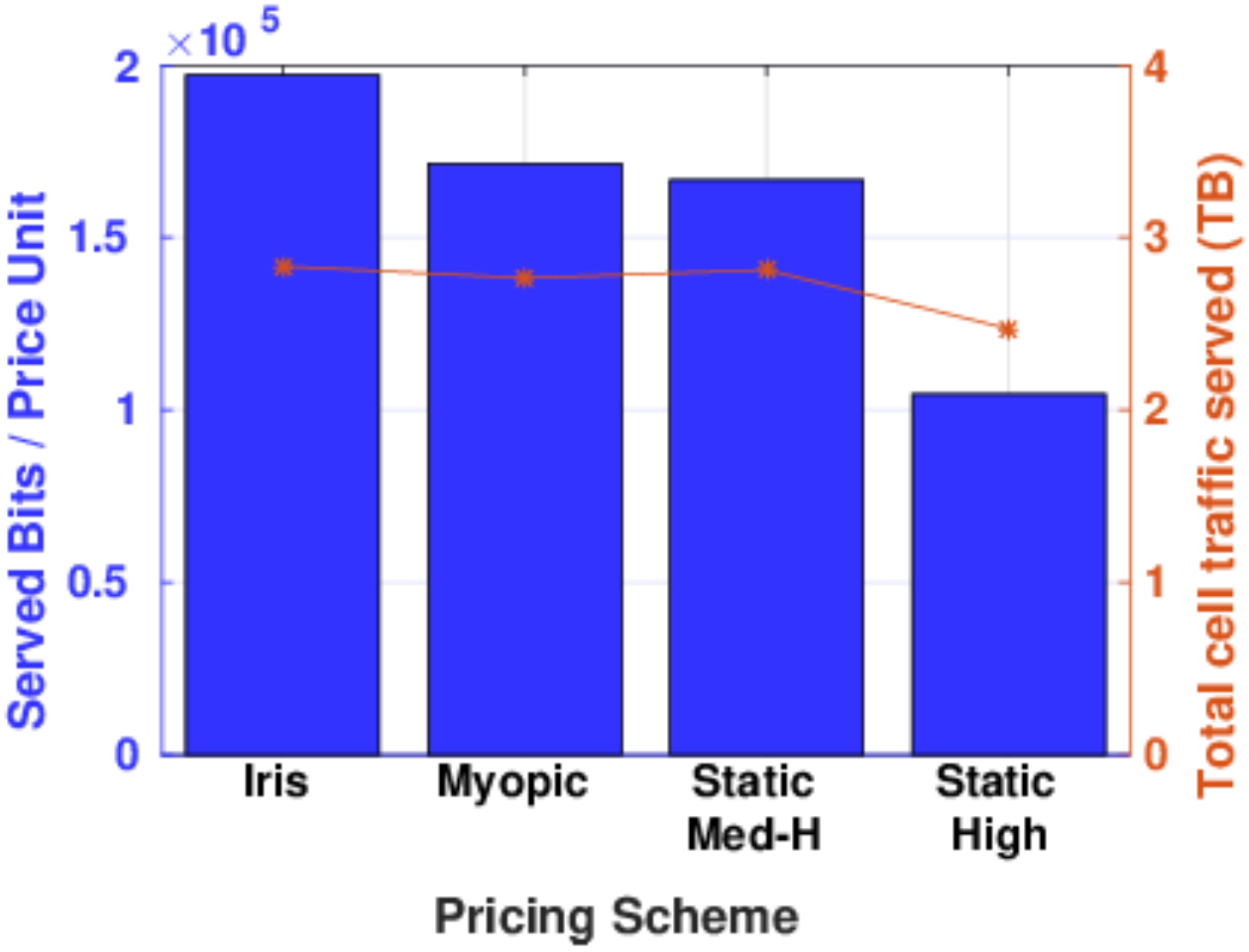}
		\caption{Total traffic served and bits per price unit.}
		\label{fig:bit_per_buck}
	\end{subfigure}
    ~~~~~~~~
    \begin{subfigure}[b]{0.45\columnwidth}  
		\centering 
		\includegraphics[width=\textwidth]{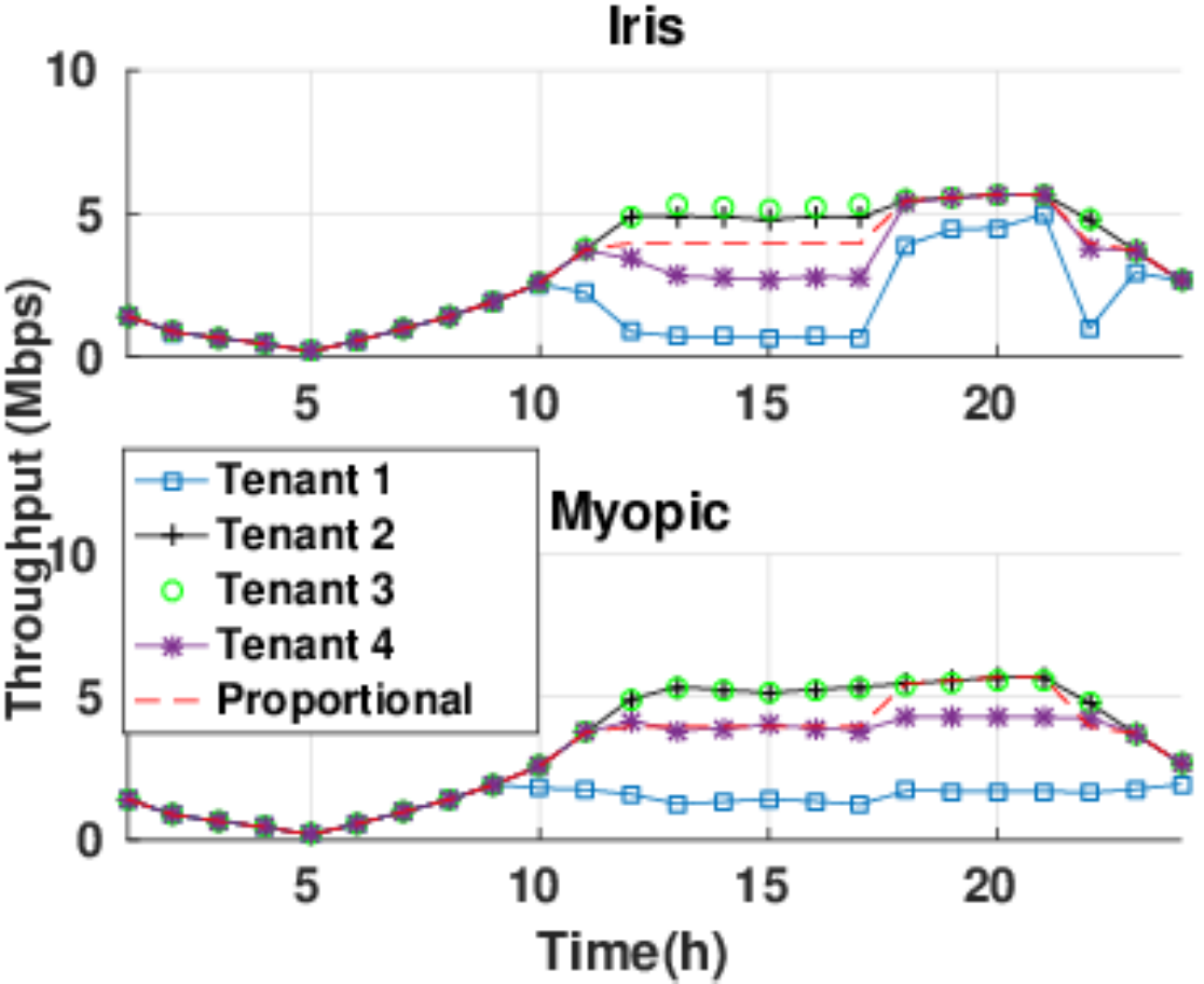}
		\caption{\iriss and myopic scheme service differentiation.}
		\label{fig:iris_comparison}
	\end{subfigure}
    \caption{Comparison of \iriss with alternative approaches.}
 	\label{fig:convergence_comparison}
 	\vspace{-2mm}
\end{figure}

In terms of the offered service, we measure the total traffic served by a cell throughout the day and calculate the average bits per price unit that the tenants bought for each pricing scheme. The results are in Fig.~\ref{fig:bit_per_buck}. We omit DNRP, DRP and the static Low and Med-L schemes, given the losses they incur to the neutral-host. As we can observe, \iriss offers the cheapest service, benefiting from the adaptiveness of its pricing scheme. Note that, although the same adaptiveness is also offered by the myopic scheme, the fact that tenants might be forced to buy unwanted resources raises the overall service cost. Another interesting observation is that, the total traffic served in the High static pricing scheme is significantly lower than that of \iris. This is because, due to the high prices, the tenants avoid buying radio resources despite the availability (evident from the fact that the other pricing schemes served more traffic with the same overall amount of resources). \looseness=-1

Finally, we compare the service differentiation offered by \iriss against the myopic scheme and the spectrum allocation policy proposed in \cite{kibria2017shared}. The latter allocates RBs to the tenants proportionally to their load, so it can be viewed as a purely load dependent but pricing agnostic scheme. For this result, the myopic scheme can act as a baseline, since the neutral-host is aware of the dis-utility functions of the tenants and thus optimally distributes the resources among them. The results appear in Fig.~\ref{fig:iris_comparison}. As we can observe, \iriss provides service differentiation among tenants, with results that are close to that of the myopic scheme. For the proportional scheme, no differentiation can be achieved (since every tenant generates the same traffic load). This can have a negative impact to the tenants' satisfaction, since the tenants that value the available spectrum the most end up getting less resources than they would like during hours of congestion (e.g., 12-6pm).\looseness=-1

\section{Discussion and Future Directions}

We believe that our work opens up a number of interesting research opportunities, which we discuss here.

\noindent \textbf{Strategic tenants.} In the current work it is assumed that tenants present a behavior that is invariant to the choices of other tenants and to their capability of affecting the price announced by the neutral-host through their actions. However, it is natural to expect that in many cases tenants could also develop strategic behavior, e.g. use their own learning agents, making resource requests that optimize their long-term benefits given the prices announced by the neutral-host. In such scenarios, we no longer have time-invariant transition rates from the point of view of any one agent (neutral-host or tenant), which can make the problem of solving the model much harder. One way to overcome this challenge could be to consider the problem in the context of a multi-agent reinforcement learning framework like~\cite{lowe2017multi}. Another approach could be to restrict the way that tenants behave and request resources, by enforcing the use of a mechanism that prohibits strategic behavior. Such a mechanism could for example restrict the frequency with which tenants can change their policy or to employ a domain specific language through which the \iriss tenant agents could express their business models and demands in a constrained way. 

\noindent \textbf{Tenant tradeoffs driving the use of the neutral-host deployment.} The focus of this work has been on the neutral-host side and on how to identify a radio resource pricing policy that can allow the neutral-host to match the available radio resource supply with the tenant demand while reaching a certain revenue target. As already mentioned, our work makes no assumptions about how the tenant spectrum acquisition behavior should be modelled, as long as the interactions of tenants with the neutral-host adhere to the dynamic pricing mechanism presented in Section~\ref{sec:multi_design}. Regardless of the method used to model the tenant behavior (e.g. utility function or something else), a very important factor that most probably should be taken into account is the revenue that the tenant is expected to make by distributing the obtained resources to its associated UEs and how this relates to the tenant's OpEx. Since different tenants can have different business models, developing a variety of \iriss tenant agent implementations that can capture the demands of those business models is a very interesting and important problem for future research.  

Another important factor that can drive the tenant behavior is that, as already mentioned at the outset, traditional operators have the option to serve users in indoor spaces either by using their own outdoor RAN infrastructure or by using the indoor neutral-host deployment after paying some fee. Due to these options that operators have, a complementary problem to the one considered here is how operators should decide whether it is preferable to use the neutral-host's infrastructure or to rely on their own. This decision could involve aspects like the level of the fee charged by the neutral-host, the number of users that would benefit from the presence of the operator in the indoor space, as well as the performance improvement that the users would experience through that. 

\noindent \textbf{Co-existence of multiple neutral-hosts in other settings.} As already explained, in the setting of this work only a single neutral-host is expected to exist (e.g. due to indoor space constraints and regulations), with its main incentive for providing its service being the improvement of the quality of experience of residents and visitors. However, when considering settings where multiple neutral-hosts could be co-located (e.g. outdoor settings) the goals of the system could change. In such settings, attracting more users and maximizing profit would also be equally significant for the neutral-host in addition to the ones like the regulation of spectrum among tenants considered in this paper. In such scenarios an alternative framework would be required (e.g. a game-theoretic framework) to drive the behavior of each individual neutral-host, considering the actions of the other neutral-hosts. 

\noindent \textbf{Spectrum management related issues.} One obvious extension of \iriss is to expand its scope to also support pooled licensed as well as unlicensed spectrum. While our system design and dynamic pricing mechanism would still form the core solution in both of these cases, modifications would also be required due to the idiosyncrasies that these scenarios present. For pooled licensed spectrum, pricing needs to additionally account for revenue sharing with MNOs contributing licensed spectrum to the pool. On the other hand, in the case of unlicensed spectrum, coexistence issues with other technologies like Wi-Fi need to be addressed (e.g. using a technology like MulteFire~\cite{alliance2017multefire}).\looseness=-1 

The dynamics of the interaction between the spectrum manager of \iriss and the external repositories for shared spectrum acquisition is another relevant topic. Deciding on the amount of spectrum to request from an external repository can be a challenging problem for the neutral-host, due to the different loads and demands presented by different small-cells, which create a requirement to draw a balance between the spectrum acquisition cost and the satisfaction of the tenants' demands. 

\noindent \textbf{Multi-RAT support.} Another interesting research topic is providing support for \iriss in multi-RAT settings. Accommodating multiple disparate radio access technologies (e.g., 5G New Radio, LTE and Wi-Fi) as part of the same neutral-host system architecture is an approach in line with the 5G vision of native multi-access with an access agnostic core network architecture. However multi-RAT support presents its own set of challenges, with the main problem being on how tenants should decide which of the available technologies to use to accommodate the needs of their users, considering that each technology presents its own pros and cons in terms of performance, cost, capacity etc.\looseness=-1

\section{Conclusions}

We have presented \iris, a system architecture for neutral-host indoor small-cells based on shared spectrum. The design of \iriss follows a C-RAN approach that allows scalable and efficient use of resources in the edge cloud while enable denser and cheaper small-cell radio infrastructure indoors. At the core of \iriss lies a novel dynamic pricing radio resource allocation mechanism for shared spectrum. This mechanism employs deep reinforcement learning to discover pricing policies that allow tenants to request shared spectrum resources on demand, ensuring the differentiation of their services based on their valuation of the spectrum, while meeting the revenue target of the neutral-host that includes recouping the costs for shared spectrum acquistion. Using our prototype implementation of \iriss developed for LTE, we have conducted extensive experimental evaluations to characterize the dynamic pricing mechanism of \iriss under different conditions, show the benefits of the \iriss approach compared to alternative approaches and examine its deployment feasibility.\looseness=-1

\section*{Acknowledgment}

The authors would like to thank Prof. George D. Stamoulis and Dr. Stefano V. Albrecht for their very helpful suggestions on improving this work.

\bibliographystyle{IEEEtran}
\bibliography{iris}

\appendix[Modelling the effect of traffic load and unit price on the behavior of the neutral-host tenants]

\subsection{Preliminaries}


Let~$d\ge0$ be the traffic load of the tenant and let $p\ge0$ be
the price charged by the neutral-host per resource block. Suppose
that the tenant requests $b\ge0$ resource blocks from the neutral-host.
This decision has a cost equal to $pb$ and creates a backlog of traffic equal to $[d-b]^+$ (using the standard notation
$[x]^+ \triangleq \maxtwo{x}{0}$).
Obviously, higher values of~$b$ reduce (or eliminate) the backlog, but also
increase the cost.

To express this trade-off, we consider dis-utility functions
$\bar{U}: \mathbb{R}^+_0\to \mathbb{R}^+_0$ of the form
\begin{equation}\label{eq:disutil-fun}
\bar{U}(b; d,p) =
 \Bigl( a\bigl([d-b]^+\bigr)^{\gamma_d}
           + \bigl(pb\bigr)^{\gamma_p}\Bigr)^{1/\gamma_p},
 \qquad a>0,\quad \gamma_d,\gamma_p\ge1.
\end{equation}
The exponents $\gamma_d$~and $\gamma_p$ in~\eqref{eq:disutil-fun} tune the
sharpness of the dissatisfaction associated with the backlog of traffic
and with the cost of the resources respectively, while the factor~$a$ expresses the relative importance of these two sources of dissatisfaction in the overall dis-utility. In the following, we will investigate the use of the dis-utility
function for determining the optimal resource allocation request
$b^*(d,p)\triangleq \argmin_{b\ge0} \bar{U}(b; d,p)$, under a
given price and traffic load.

With respect to the units of the various quantities involved, and apart from the
two exponents $\gamma_d$ and~$\gamma_p$, which are dimensionless,
$d$ and~$b$ are expressed in units of~$\units{resource~block}$.
The price~$p$ is in units of $\units{cost}/\units{resource~block}$,
while the factor~$a$ is in units of
$\units{cost}^{\gamma_p}/\units{resource~block}^{\gamma_d}$.
Finally, since the parenthesized sum in~\eqref{eq:disutil-fun} is raised to
the power of~$1/\gamma_p$, the values of the dis-utility function $\bar{U}$
are expressed in units of~$\units{cost}$.

This last feature is worthwhile because, apart from the use of the dis-utility
function for determining~$b^*$, the function is useful also for quantifying the
dis-utility experienced by multiple tenants, whose
dis-utility functions may employ different values for the parameters $\gamma_d$,
$\gamma_p$ and~$a$. Regardless of such differences, all dis-utilities
will be expressible in the same unit of~$\units{cost}$, bearing the same
interpretation for all tenants and being directly comparable. Additionally,
it is possible to introduce a notion of `overall dis-utility', calculated
as the sum of dis-utilities over all tenants.\looseness=-1

\subsection{Structural properties}

Since $(\cdot)^{\gamma_p}$ is strictly increasing, the minima
of $\bar{U}(\cdot\,; d, p)$
and of $U(\cdot\,; d, p)\triangleq \bar{U}(\cdot; d, p)^{\gamma_p}$, coincide.
Thus, we may work with the simpler function~$U$, equal to the
parenthesized sum in~\eqref{eq:disutil-fun}.

By construction, $U$~is continuous and convex (because $\gamma_d,\gamma_p\ge1$)
throughout its domain.
In fact, $U$~is strictly convex (and its derivative strictly increasing), 
except when $\gamma_d=\gamma_p=1$, in which case it is piecewise linear.

In view of~\eqref{eq:disutil-fun}, $U$~is continuously differentiable in
$[0,d)\cup(d,+\infty)$, with
\begin{equation}\label{eq:disutil-deriv}
U'(b; d, p) = \begin{cases}
-a\gamma_d (d-b)^{\gamma_d-1} + \gamma_p p^{\gamma_p} b^{\gamma_p-1},& 0\le b<d,\\
\gamma_p p^{\gamma_p} b^{\gamma_p-1},                                & b>d.
\end{cases}
\end{equation}
The derivative in~\eqref{eq:disutil-deriv} is continuous also at $b=d$ and
$U'(d)$ exists unless $\gamma_d=1$, in which case $U'(d^-) < U'(d^+)$.

By the second branch in~\eqref{eq:disutil-deriv}, $U$ is increasing for $b>d$,
so the infimum of the function occurs within the closed and bounded
interval~$[0,d]$ and, by continuity, there exists a minimal point
$b^*\in[0,d]$. Furthermore, by convexity, this minimal point is unique (except
perhaps for the non-strictly convex case $\gamma_d=\gamma_p=1$ when the minimum
may be attained for all points of an interval within~$[0,d]$).

\subsection{The optimal resource allocation request~$b^*(d,p)$}
We now express~$b^*$ as a function of the given traffic load and price.
As we will see, the shape of this function depends on the values of the
exponents $\gamma_d$ and~$\gamma_p$.

\subsubsection{The case $\gamma_d=\gamma_p=1$ -- extreme behavior}\label{sec:extreme}
By~\eqref{eq:disutil-deriv}, $U$~is decreasing
throughout~$[0,d]$ when~$p<a$, increasing when~$p>a$, and constant when $p=a$.
Thus,
\begin{equation}\label{eq:extreme-b}
b^*(d, p) = \begin{cases}
  d,                        & p<a,\\
  \text{any $b\in[0,d]$},   & p=a,\\
  0,                        & p>a.
\end{cases}
\end{equation}
The form~\eqref{eq:extreme-b} signifies ``extreme'' behavior. The factor~$a$
fixes a price threshold and the tenant either makes a resource request equal to the traffic load when the price is below the threshold, or backlogs all its traffic when the price exceeds the threshold.

\subsubsection{The case $\gamma_d=1$, $\gamma_p>1$ -- cost saving tendency;
            limiting allocations below a price-dependent threshold}
\label{sec:low-dem}

By~\eqref{eq:disutil-deriv}, if $U'(d^-)\le0$, i.e.,
if $a \ge \gamma_p p^{\gamma_p} d^{\gamma_p-1}$ then
$U$~is decreasing throughout~$[0,d]$ and $b^*=d$. Otherwise, the minimization
occurs at the unique solution of $U'(b)=0$. Putting these facts together,
\begin{equation}\label{eq:lowdem-b}
b^*(d, p) =
 \mintwo{%
    \Bigl(\frac{a}{\gamma_p p^{\gamma_p}}\Bigr)^{\frac{1}{\gamma_p-1}}%
  }{d}.
\end{equation}
This result may be interpreted as follows: Fixing a price
determines a traffic load threshold~$d_0$, equal to the first argument of the
$\min$-operator within~\eqref{eq:lowdem-b}. For traffic loads lower than this threshold the tenant requests resource blocks equal to the load; for higher ones, resource blocks for a load of~$d_0$ are requested and the remaining part of the traffic is backlogged.

An alternative interpretation may also be given, by first fixing the
traffic load~$d$ and then varying~$p$. Along this interpretation, as prices
increase, starting from an initially low level close to~0, the traffic is
fully covered. This occurs up to a load-dependent price threshold
(obtained by equating the two arguments of the $\min$-operator
in~\eqref{eq:lowdem-b} and then solving for $p$), while higher prices
introduce backlog. In this regime, the processed part of the traffic gradually diminishes as $p\to\infty$.\looseness=-1

It may be seen that as $\gamma_p\downto1$, the case considered in this section
tends to the extreme case in Section~\ref{sec:extreme} and \eqref{eq:lowdem-b}
collapses to~\eqref{eq:extreme-b}.

\subsubsection{The case $\gamma_d>1$, $\gamma_p=1$ -- limiting backlogs below a
            price-dependent threshold}
\label{sec:hi-dem}

In view of~\eqref{eq:disutil-deriv},
when $U'(0)\ge0$, i.e., when $p\ge a\gamma_d d^{\gamma_d-1}$,
the function~$U$ is increasing throughout~$[0,d]$ and $b^*=0$. Otherwise,
the minimization occurs at the unique solution of $U'(b)=0$. Overall,
\begin{equation}\label{eq:hidem-b}
b^*(d,p) =
  \left[ d - \Bigl(\frac{p}{a\gamma_d}\Bigr)^{\frac{1}{\gamma_d-1}} \right]^+.
\end{equation}
According to this result, fixing a price again determines a traffic load
threshold~$d_0$, now equal to the term subtracted from~$d$
in~\eqref{eq:hidem-b}. For traffic loads higher than the threshold the tenant requests resource blocks that will only partially cover its load, creating a backlog of remaining traffic equal to~$d_0$. Loads
lower than the threshold are entirely backlogged. This behavior tends to favor
higher traffic loads over lower ones.

For the alternative interpretation related to fixed traffic loads and varying prices, it may be seen that as prices increase, again starting from an initially low level close to~0, the unprocessed part of the traffic gradually increases. Past a load-dependent price (obtained by equating the two terms subtracted in~\eqref{eq:hidem-b} and solving for~$p$), the entire traffic load is backlogged.

Again, it may be seen that as $\gamma_d\downto1$, the case considered here
tends to the extreme case in Section~\ref{sec:extreme} and \eqref{eq:hidem-b}
collapses to~\eqref{eq:extreme-b}.
%

\subsubsection{The case $\gamma_d>1$, $\gamma_p>1$ and the balanced sub-case~$\gamma_d=\gamma_p=\gamma>1$}
\label{sec:balanced}
Now, $U'(0)<0$, and $U'(d)>0$, so $b^*$~lies in the interior of~$[0,d]$ and
is determined as the unique solution of $U'(b)=0$, equivalently the unique
solution of the non-linear equation (in~$b$)
\begin{equation}\label{eq:gencase}
\Bigl(
\frac{\gamma_p p^{\gamma_p}}{\gamma_d a}
\Bigr)^{\frac{1}{\gamma_d-1}} b^{\frac{\gamma_p-1}{\gamma_d-1}}
  + b = d.
\end{equation}
Generally, this equation must be solved numerically, but for
$\gamma_d-1 = 2(\gamma_p-1)$ (closer in spirit to the case of
Section~\ref{sec:hi-dem}) and for $\gamma_d-1 = (\gamma_p-1)/2$
(closer in spirit to the case of Section~\ref{sec:low-dem}) the
equation reduces to a quadratic leading to a closed form solution.

Again, it may be seen that by keeping $\gamma_p$ fixed and letting
$\gamma_d\downto1$
the case in this section tends to the case in Section~\ref{sec:low-dem}.
Similarly, by keeping $\gamma_d$ fixed and letting
$\gamma_p\downto1$, this case reduces to the case of
Section~\ref{sec:hi-dem}. Finally, letting both exponents tend to unity leads
to the case in Section~\ref{sec:extreme}.\looseness=-1

When the exponents are equal and greater than unity, \eqref{eq:gencase}~collapses
to a first order linear equation with solution
\begin{equation}\label{eq:balanced}
b^*(d, p) = \frac{d}{1 + (p^\gamma/a)^{\frac{1}{\gamma-1}}}.
\end{equation}
It is seen that all levels of traffic loads are treated uniformly,
with the tenant requesting resource blocks that will neither fully cover nor entirely backlog the existing traffic load. Instead, the price
determines the fraction of the traffic to be processed. As $\gamma\downto1$ this solution adopts the sharp characteristics
of the case in Section~\ref{sec:extreme}.

\subsection{Parametrizing the dis-utility function}
Here we consider values for the parameters of the dis-utility
function~$\bar{U}$, to tailor the function to a particular tenant's
behavior and create the tenant profiles presented in Table~\ref{tab:tenant_profiles} of Section~\ref{sec:eval_setup}.  We address the cases in Sections \ref{sec:low-dem}, \ref{sec:hi-dem} and~\ref{sec:balanced}. In each of these, one needs to determine values for an exponent and for the
factor~$a$. The value of the exponent is chosen first, to determine the
``sharpness'' of the function's response. Then, to determine the factor~$a$ one proceeds as follows:
\begin{itemize}[leftmargin=*]
\item For the ``cost saving'' case in Section~\ref{sec:low-dem}, one specifies a
traffic load threshold~$d_0$ and a price threshold~$p_0$. These determine the
value of~$a$,
so that for prices and loads not greater than the thresholds the traffic load of a tenant is fully processed. By equating the two arguments of the $\min$-operator
in~\eqref{eq:lowdem-b}, with the price and load therein set equal to the
thresholds, the appropriate value of the factor is seen to be
\begin{equation}\label{cost-a}
a = \gamma_p p_0^{\gamma_p} d_0^{\gamma_p-1}.
\end{equation}

Profiles 1 and 2 of Table~\ref{tab:tenant_profiles} were based on this ``cost saving'' case, with $\gamma_p=2$ (and $\gamma_d=1$). For the first profile, a traffic load threshold~$d_0=1750$ and a price threshold~$p_0=100$ were used, leading to a value of~$a=3.5\times10^8$ and a ``best effort'' profile type, in which the tenant is willing to fully cover its load for very low prices (recall that $p_{\max}=2500\gg p_0$) and only a small part of the load otherwise (just to maintain network presence). For the second profile, a traffic load threshold~$d_0=1000$ and a price threshold~$p_0=1000$ were used ($a=2\times10^9$). This leads to a ``price-driven'' profile that is similar to the first one, with the difference that the tenant is willing to buy more resources for low to medium prices compared to the best-effort case (not just focusing on maintaining network presence). 
\item For the ``bounded backlog'' case in Section~\ref{sec:hi-dem}, one again
specifies a traffic load threshold~$d_0$ and a price threshold~$p_0$. These determine the value of~$a$, so that for prices greater than~$p_0$ and loads lower than~$d_0$ the traffic is backlogged entirely. By equating the two terms subtracted
in~\eqref{eq:hidem-b}, with the price and load therein set equal to the
thresholds, the appropriate value of the factor is seen to be
\begin{equation}\label{backlog-a}
a = \frac{p_0}{\gamma_d d_0^{\gamma_d-1}}.
\end{equation}

The third profile of Table~\ref{tab:tenant_profiles} was based on this analysis, with $\gamma_d=2$ (and $\gamma_p=1$). The traffic load threshold was set to~$d_0=6000$ and the price threshold to~$p_0=2436$ (i.e., close to~$p_{\max}$), leading to a value of $a=0.203$. This profile corresponds to a ``demand-driven'' tenant, who is willing to buy large amounts of resources regardless of the price when the traffic load is high and to queue its traffic until the load increases enough to buy in bulk in other times.
\item Finally, for the balanced sub-case in Section~\ref{sec:balanced}, one
must specify a price threshold~$p_0$ and the corresponding fraction~$\omega_0\in(0,1)$ of the traffic load that will be processed. Then, in view of~\eqref{eq:balanced}, the appropriate value of the factor is seen to be
\begin{equation}\label{balanced-a}
a = p_0^\gamma \Bigl(\frac{\omega_0}{1-\omega_0}\Bigr)^{\gamma-1}.
\end{equation}
Correspondingly, the last profile of Table~\ref{tab:tenant_profiles} was set to represent a ``medium'' QoS level type of tenant with $\gamma=2$, $p_0=600$ and $\omega_0\approx0.25$, yielding $a=1.1\times10^5$.

\end{itemize}

\end{document}